\documentclass[hyper,11pt,letterpaper]{JHEP3}

\usepackage{cite, graphicx, subfigure, amsmath,amssymb,calrsfs,mathrsfs}

\newcommand{\mc}[1]{\mathscr{#1}}
\newcommand{\rv}[1]{\boldsymbol{\mathsf{#1}}}
\newcommand{\ra}{\rightarrow}

\title{Continuous-Discrete Path Integral Filtering}

\author{Bhashyam Balaji,\\
Radar Systems Section,\\
Defence Research and Development Canada, Ottawa,\\
3701 Carling Avenue, \\
Ottawa ON K1A 0Z4 Canada\\
Email: Bhashyam.Balaji@drdc-rddc.gc.ca}

\maketitle

\abstract{
	A summary of the relationship between the Langevin equation, Fokker-Planck-Kolmogorov forward equation (FPKfe) and the Feynman path integral descriptions of stochastic processes relevant for the solution of the continuous-discrete filtering problem is provided in this paper. The practical utility of the path integral formula is demonstrated via some nontrivial examples. Specifically, it is shown that the simplest approximation of the path integral formula for the fundamental solution of the FPKfe can be applied to solve nonlinear continuous-discrete filtering problems quite accurately. The Dirac-Feynman path integral filtering algorithm is quite simple, and is suitable for real-time implementation. }


\begin{document}

\section{Introduction}
The following continuous-discrete filtering problem often arises in practice. The time evolution of the state, or signal of interest, is well-described by a continuous-time stochastic process. However, the state process is not directly observable, i.e., the state process is a hidden continuous-time Markov process. Instead, what is measured is a related discrete-time stochastic process termed the measurement process. The continuous-discrete filtering problem is to estimate the state of the system, given the measurements \cite{A.H.Jazwinski1970}.

When the state and measurement processes are linear, excellent performance is often obtained using the Kalman filter \cite{R.E.Kalman1960,R.E.KalmanR.S.Bucy1961}. However, the Kalman filter merely propagates the conditional mean and covariance, so is not a universally optimal filter and is inadequate for problems with non-Gaussian characteristics (e.g., multi-modal). When the state and/or measurement processes are nonlinear, a (non-unique) linearization of the problem leads to an extended Kalman filter.If the nonlinearity is benign, it is still very effective. However, for the general case, it cannot provide a robust solution.

The complete solution of the filtering problem is the conditional probability density function of the state given the observations. It is complete in the Bayesian sense, i.e., it contains all the probabilistic information about the state process that is in the measurements and the initial condition. The solution is termed universal if the initial distribution can be arbitrary. From the conditional probability density, one can compute quantities optimal under various criteria. For instance, the conditional mean is the least mean-squares estimate.

The solution of the continuous-discrete filtering problem requires the solution of a linear, parabolic, partial differential equation (PDE) termed the Fokker-Planck-Kolmogorov forward equation (FPKfe). There are three main techniques to solve the FPKfe type of equations, namely, finite difference methods \cite{Thom'ee90,Marchuk90}, spectral methods \cite{C.CanutoandM.Y.HussainiandA.QuarteroniandT.A.ZangJr.2006}, and finite/spectral element methods \cite{V.Thom'ee1997}. However, numerical solution of PDEs is not straightforward. For example, the error in a na\" ive discretization may not vanish as the grid size is reduced, i.e., it may not be convergent. Another possibility is that the method may not be consistent, i.e., it may tend to a different PDE in the limit that the discretization spacing vanishes. Furthermore, the numerical method may be unstable, or there may be severe time step size restrictions.

The fundamental solution of the FPKfe can be represented in terms of a Feynman path integral \cite{R.P.FeynmanandA.R.Hibbs1965}. The path integral formula can be derived directly from the Langevin equation. A textbook discussion for derivation of the path integral representation of the fundamental solution of the FPKfe corresponding to the Langevin equations for additive and multiplicative noise cases can be found in \cite{JeanZinn-Justin2002} and \cite{F.LangoucheandD.RoekaertsandE.Tirapegui1982}. These results have been simplified and generalized for the general case in \cite{PAPER1} and \cite{PAPER3}. In this paper, it is demonstrated that the simplest approximate path integral formulae lead to very accurate solution of the nonlinear continuous-discrete filtering problem.

In Section \ref{sec:RevCDFilt}, the basic concepts of continuous-discrete filtering theory is reviewed. In Section \ref{sec:PIFormulas}, the path integral formulae for the case of additive and multiplicative noise cases are summarized. In Section \ref{sec:PIFiltering}, an elementary solution of the continuous-discrete filtering problem is presented that is based on the path integral formulae. Some examples illustrating the path integral filtering is presented in the following section. Some remarks on practical implementational aspects of path integral filtering is presented in Section \ref{sec:AddlRemarks}. The appendix summarizes the path integral results derived in \cite{PAPER1} and \cite{PAPER3}. 

\section{Review of Continuous-Discrete Filtering Theory}\label{sec:RevCDFilt}
\subsection{Langevin Equation and the FPKfe}

The general continuous-time state model is described by the following stochastic differential equation (SDE):
\begin{align}\label{eq:LangevinEquation}
	d\rv{x}(t)=f(\rv{x}(t),t)dt+e(\rv{x}(t),t)d\rv{v}(t),\qquad \rv{x}(t_0)=x_0.
\end{align}
Here $\rv{x}(t)$ and $f(\rv{x}(t), t)$ are $n-$dimensional column vectors, the diffusion vielbein $e(\rv{x}(t), t)$ is an $n\times p_e$ matrix and $\rv{v}(t)$ is a pe¿dimensional column vector. The noise process $\rv{v}(t)$ is assumed to be Brownian with covariance $Q(t)$ and the quantity $g\equiv eQe^T$ is termed the diffusion matrix. All functions are assumed to be sufficiently smooth. Equation \ref{eq:LangevinEquation} is also referred to as the Langevin equation. It is interpreted in the It\^o sense (see Appendix \ref{sec:AppendixI}). Throughout the paper, bold symbols refer to the stochastic processes while the corresponding plain symbol refers to a sample of the process.

Let $\sigma_0(x)$ be the initial probability distribution of the state process. Then, the evolution of the probability distribution of the state process described the Langevin equation, $p(t, x)$, is described by the FPKfe, i.e.,
\begin{align}
        {\left\lbrace\begin{aligned}    
                \frac{\partial p}{\partial t}(t,x)&=-\sum_{i=1}^n\frac{\partial}{\partial x_i}\left[ f_i(x,t)p(t,x) \right]+\frac{1}{2}\sum_{i,j=1}^n\frac{\partial^2}{\partial x_i\partial x_j}\left[g_{ij}(t,x)p(t,x)\right], \\ 
        p(t_0,x)&=\sigma_0(x).
\end{aligned}
\right.}
\end{align}

\subsection{Fundamental Solution of the FPKfe}
The solution of the FPKfe can be written as an integral equation. To see this, first note that the complete information is in the transition probability density which also satisfies the FPKfe except with a $\delta-$function initial condition. Specifically, let $t'' > t'$, and consider the following
PDE:
\begin{align}
        {\left\lbrace\begin{aligned}    
                \frac{\partial P}{\partial t}(t,x|t',x')&=-\sum_{i=1}^n\frac{\partial}{\partial x_i}\left[ f_i(x,t)P(t,x|t',x') \right]+\frac{1}{2}\sum_{i,j=1}^n\frac{\partial^2}{\partial x_i\partial x_j}\left[g_{ij}(t,x)P(t,x|t',x'))\right], \\ 
        P(t',x''|t',x')&=\delta^n(x''-x').\end{aligned}
\right.}
\end{align}
Such a solution, i.e., $P(t,x|t',x')$, is also known as the fundamental solution of the FPKfe. From the fundamental solution one can compute the probability at a later time for an arbitrary initial condition as follows:
\begin{align}\label{eq:EvolFPKfeSolnIntegral}
	p(t'',x'')=\int P(t'',x''|t',x')p(t',x')\left\{ d^nx' \right\}.
\end{align}
In this paper, all integrals are assumed to be from $-\infty$ to $+\infty$, unless otherwise specified. Therefore, in order to solve the FPKfe it is sufficient to solve for the transition probability density $P(t, x|t', x')$. Note that this solution is universal in the sense that the initial distribution can be arbitrary.

\subsection{Continuous-Discrete Filtering}

In this paper, it is assumed that the measurement model is described by the following discrete-time stochastic process
\begin{align}
	\rv{y}(t_k)=h(\rv{x}(t_k),\rv{w}(t_k),t_k),\qquad k=1,2,\ldots,\qquad t_k>t_0,
\end{align}
where $y(t)\in \mathbb{R}^{m\times1}, h\in \mathbb{R}^{m\times1}$, and the noise process $\rv{w}(t)$ is assumed to be a white noise process.  Note that $\rv{y}(t_0)=0$. It is assumed that $p(y(t_k)|x(t_k))$ is known.

Then, the universal continuous-discrete filtering problem can be solved as follows. Let the initial distribution be $\sigma_0(x)$ and let the measurements be collected at time instants $t_1, t_2,\ldots, t_k,\ldots$. Let $p(t_{k-1}|Y(t_{k-1}))$ be the conditional probability density at time $t_{k-1}$, where  $Y(\tau) = \{y(t_l) : t_0 < t_l\le \tau\}$. Then the conditional probability density at time $t_k$, after incorporating the measurement $y(t_k)$, is obtained via the prediction and correction steps:
\begin{align}
{\left\lbrace\begin{aligned} 
	p(t_k,x|Y(t_{k-1}))&=\int P(t_k,x|t_{k-1},x_{_k-1})p(t_{_k-1},x_{_k-1}|Y(t_{_k-1}))\left\{ d^nx_{k-1} \right\},\qquad\text{(Prediction Step)},\\ 
	p(t_k,x|Y(t_k))&=\frac{p(y(t_k)|x)p(t_k,x|Y(t_{k-1}))}{\int p(y(t_k)|\xi)p(t_k,\xi|Y(t_{k-1}))\left\{ d^n\xi \right\}},\qquad\text{(Correction Step)}. 
\end{aligned}\right.}
\end{align}
Often (as in this paper), the measurement model is described by an additive Gaussian noise model, i.e.,
\begin{align}
	\rv{y}(t_k)=h(\rv{x}(t_k),t_k)+\rv{w}(t_k),\qquad k=1,2,\ldots,\qquad t_k>t_0,
\end{align}
with $\rv{w}(t)\sim N(0,R(t))$, i.e.,
\begin{align}
        p(y(t_k)|x)=\frac{1}{\left( (2\pi)^m\det R(t_k) \right)^{1/2}}\exp\left\{ -\frac{1}{2}(y(t_k)-h(x(t_k),t_k))^T(R
(t_k))^{-1}(y(t_k)-h(x(t_k),t_k)) \right\},
\end{align}

Observe that, as in the PDE formulations, one may use a convenient set of basis functions. Then, the evolution of each of the basis functions under the FPKfe follows from Equation \ref{eq:EvolFPKfeSolnIntegral}. Since the basis functions are independent of measurements, the computation may be performed off-line. Finally, note that this solution of the filtering problem is universal. In conclusion, the determination of the fundamental solution of the FPKfe is equivalent to the solution of the universal optimal nonlinear filtering problem. A solution for the time independent case with orthogonal diffusion matrix in terms of ordinary integrals was presented in \cite{S.-T.YauS.S.-TYau1996}. However, the integrand is complicated and not easily implementable in practice. In the next section,the fundamental solution for the general case in terms of path integrals is summarized. It is shown that it leads to formulae that are simple to implement.

\section{Path Integral Formulas}\label{sec:PIFormulas}
In this section, path integral formulae derived in \cite{PAPER1} and \cite{PAPER3} are summarized. It is assumed that $t'' > t'$. Details on the formulae are summarized in Appendix \ref{sec:AppendixI}.
\subsection{Additive Noise}
When the diffusion vielbein is independent of the state,  i.e.,
\begin{align}
	d\rv{x}(t)=f(\rv{x}(t),t)dt+e(t)d\rv{v}(t),
\end{align}
where all quantities are as defined in Section II-C, the noise is said to be additive. The path integral formula for the transition probability density is given by
\begin{align}
	P(t'',x''|t',x')=\int_{x(t')=x'}^{x(t'')=x''}\left[ \mc{D}x(t) \right]\exp\left( -\int_{t'}^{t''}dt L^{(r)}(t,x,\dot{x}) \right),
\end{align}
where the Lagrangian $L^{(r)}(t,x,\dot{x})$ is defined as
\begin{align}\label{eq:LagrangianAdd}
	L^{(r)}(t,x,\dot{x})=\frac{1}{2}\sum_{i=1}^n\left( \dot{x}_i-f_i(x^{(r)}(t),t) \right)g^{-1}_{ij}(t)\left( \dot{x}_j-f_j(x^{(r)}(t),t) \right)+r\sum_{i=1}^n\frac{\partial f_i}{\partial x_i}(x,t),
\end{align}
and
\begin{align}
	g_{ij}(t)=\sum_{a,b=1}^{p_e}e_{ia}(t)Q_{ab}(t)e_{bj}(t),
\end{align}
and 
\begin{align}
	\left[ \mc{D}x(t) \right]=\frac{1}{\sqrt{(2\pi\epsilon)^n\det g(t')}}\lim_{N\ra\infty}\prod_{k=1}^N\frac{d^nx(t'+k\epsilon)}{\sqrt{(2\pi\epsilon)^n\det g(t'+k\epsilon)}}.
\end{align}
Here, $r\in[0,1]$ specifies the discretization of the SDE; see Appendix \ref{sec:AppendixI} for details. The quantity $S(t'',x')=\int_{t'}^{t''}L^{(r)}(t,x,\dot{x})dt$ is referred to as the action.

\subsection{Multiplicative Noise}
The state model for the general case is given by
\begin{align}
	d\rv{x}(t)=f(\rv{x}(t),t)dt+e(\rv{x}(t),t)d\rv{v}(t).
\end{align}
As discussed in more detail in Appendix \ref{sec:AppendixI}, there is ambiguity in the definition of this SDE which is due to the fact that $d\rv{v}(t)\approx O(\sqrt{dt})$. The path integral formula for the general discretization is complicated and summarized in Appendix \ref{sec:AppendixI}. In the simplest It\^o case, it reduces to
\begin{align}
	P(t'',x''|t',x')=\int_{x(t')=x'}^{x(t'')=x''}\left[ \mc{D}x(t) \right]\exp\left( -\int_{t'}^{t''}dt L^{(r,0)}(t,x,\dot{x}) \right),
\end{align}
where the Lagrangian $L^{(r,0)}(t, x, \dot{x} )$ is defined as
\begin{align}\label{eq:LagrangianMultIto}
	L^{(r,0)}(t,x,\dot{x})=\frac{1}{2}\sum_{i=1}^n\left( \dot{x}_i-f_i(x^{(r)}(t),t) \right)g^{-1}_{ij}(x^{(0)}(t),t)\left( \dot{x}_j-f_j(x^{(r)}(t),t) \right)+r\sum_{i=1}^n\frac{\partial f_i}{\partial x_i}(x^{(r)},t),
\end{align}
and
\begin{align}
	g_{ij}(x^{(0)}(t),t)=\sum_{a,b=1}^{p_e}e_{ia}(x^{(0)}(t),t)Q_{ab}(t)e_{bj}(x^{(0)}(t),t).
\end{align}
A nice feature of the It\^o interpretation is that the formula is the same as that for the simpler additive noise case (with some obvious changes). 

Note that it is always possible to convert from a SDE defined in any sense (say, Stratanovich or s = 0) to the corresponding It¿o SDE. Therefore, this can be considered to be the result for the general case.

\section{Dirac-Feynman Path Integral Filtering}\label{sec:PIFiltering}

The path integral is formally defined as the $N\ra\infty$ limit of a $N$ multi-dimensional integrals and yields the correct answer for arbitrary time step size. In this section, an algorithm for continuous-discrete filtering using the simplest approximation to the path integral formula, termed the Dirac-Feynman approximation, is derived.

\subsection{Dirac-Feynman Approximation}
Consider first the additive noise case. When the time step $\epsilon\equiv t''-t'$ is infinitesimal, the path integral is given by
\begin{align}
	P(t'+\epsilon,x''|t',x')=\frac{1}{\sqrt{(2\pi\epsilon)^n\det g(t')}}\exp\left[ -\epsilon L^{(r)}(t,x',x'',(x''-x')/\epsilon) \right],
\end{align}
where the Lagrangian is
\begin{align}
	\frac{1}{2}\sum_{i,j=1}^n&\left[ \frac{x_i''-x_i'}{\epsilon}-f_i(x'+r(x''-x'),t) \right]g_{ij}^{-1}(t')\left[ \frac{x_j''-x_j'}{\epsilon}-f_j(x'+r(x''-x'),t) \right]\\ \nonumber
	&\qquad+r\sum_{i=1}^n\frac{\partial f_i}{\partial x_i}(x'+r(x''-x'),t).
\end{align}
This leads to a natural approximation for the path integral for ¿small¿ time steps:
\begin{align}
	P(t'',x''|t',x')=\frac{1}{\sqrt{(2\pi(t''-t'))^n\det g(t')}}\exp\left[ -\epsilon L^{(r)}(t,x',(x''-x')/(t''-t')) \right].
\end{align}
A special case is the one-step pre-point approximate formula
\begin{align}
	P(t'',x''|t',x')&=\frac{1}{\sqrt{(2\pi(t''-t'))^n\det g(t')}}\\ \nonumber
	&\qquad\exp\left( -\frac{(t''-t')}{2}\sum_{i,j=1}^n\left[ \frac{(x_i''-x_i')}{(t''-t')}-f_i(x',t') \right]g_{ij}^{-1}(t')\left[ \frac{(x_j''-x_j')}{(t''-t')}-f_j(x',t') \right] \right).
\end{align}
The one-step symmetric approximate path integral formula for the transition probability amplitude (as originally used by Feynman in quantum mechanics \cite{R.P.FeynmanandA.R.Hibbs1965}) is
\begin{align}\label{eq:DiscApproxAdditive}
&P(t'',x''|t',x')=\frac{1}{\sqrt{(2\pi(t''-t'))^n\det g(\bar{t})}}\\ \nonumber
	&\qquad\times\exp\left( -\frac{(t''-t')}{2}\sum_{i,j=1}^n\left[ \frac{(x_i''-x_i')}{(t''-t')}-f_i(\bar{x},\bar{t}) \right]g_{ij}^{-1}(\bar{t}) \left[ \frac{(x_j''-x_j')}{(t''-t')}-f_j(\bar{x},\bar{t}) \right]-\frac{(t''-t')}{2}\sum_{i=1}^n\frac{\partial f_i}{\partial x_i}(\bar{x},\bar{t})\right),
\end{align}
where $\bar{x}=\frac{1}{2}(x''+x')$ and $\bar{t}=\frac{1}{2}(t'+t'')$. Note that for the explicit time-dependent case the time has also been symmetrized in the hope that it will give a more accurate result. Of course, for small time steps and if the time dependence is benign, the error in using this or the end points is small.

Similarly, for the multiplicative noise case in the It\^o interpretation/discretization of the state SDE the following approximate formula results:
\begin{align}
	P(t'',x''|t',x')=\frac{1}{\sqrt{(2\pi(t''-t'))^n\det g(t')}}\exp\left[ -(t''-t') L^{(r,0)}(t,x',(x''-x')/(t''-t')) \right],
\end{align}
where the Lagrangian $L^{(r,0)}(t,x',x'',(x''-x')/(t''-t'))$ is given by
\begin{align}
	L^{(r,0)}&(t,x',x'',(x''-x')/(t''-t'))=\\ \nonumber
	&\frac{1}{2}\sum_{i,j=1}^n\left( \frac{(x_i''-x_i')}{(t''-t')}-f_i(x'+r(x''-x'),t') \right)\left( \sum_{a,b=1}^{p_e}e_{ia}(x',t')Q_{ab}(t')e_{jb}(x',t') \right)^{-1}\\ \nonumber
	&\left( \frac{(x_j''-x_j')}{(t''-t')}-f_j(x'+r(x''-x'),t') \right)+r\sum_{i=1}^n\frac{\partial f_i}{\partial x_i}(x'+r(x''-x'),t).
\end{align}
For the multiplicative noise case, the simplest one-step approximation is the pre-point discretization where $r = s = 0$:
\begin{align}\label{eq:DiscApproxMult}
	P(t'',x''|t',x')&=\frac{1}{\sqrt{(2\pi(t''-t'))^n\det g(x',t')}}\\ \nonumber
	&\times\exp\left[ -\frac{(t''-t')}{2}\left( \frac{(x_i''-x_i')}{(t''-t')}-f_i(x',t') \right)g^{-1}_{ij}(x',t') \left( \frac{(x_j''-x_j')}{(t''-t')}-f_j(x',t') \right)\right].
\end{align}
Since $s = 0$, this means that we are using the It\^o interpretation of the state model Langevin equation. When $r=1/2$, it is termed the Feynman convention, while $s=1/2$ corresponds to the Stratanovich interpretation.

\subsection{The Dirac-Feynman Algorithm}

The one-step formulae discussed in the previous section lead to the simplest path integral filtering algorithm, termed the Dirac-Feynman (DF) algorithm. 
The steps for DF algorithm may be summarized as follows:
\begin{enumerate}
	\item From the state model, obtain the expression for the Lagrangian. Specifically,
		\begin{itemize}
			\item For the additive noise case, the Lagrangian is given by Equation \ref{eq:LagrangianAdd};
			\item For the multiplicative noise case with It\^o discretization the Lagrangian is given by Equation \ref{eq:LagrangianMultIto}, while for the general discretization the action is given in Appendix \ref{sec:AppendixI}.
		\end{itemize}
	\item Determine a one-step discretized Lagrangian that depends on $r\in [0, 1]$ (and $s\in[0, 1]$ for the multiplicative noise). The usual choice is $r = 1/2$.
	\item Compute the transition probability density $P(t'', x''|t', x')$ using the appropriate formula (e.g., Equations \ref{eq:DiscApproxAdditive} or \ref{eq:DiscApproxMult}). The grid spacing should be such that the transition probability tensor is adequately sampled, as discussed below. 
	\item At time $t_k$
		\begin{enumerate}
			\item The prediction step is accomplished by
				\begin{align}
					p(t_k|Y(t_{k-1}))=\int P(t_k,x|t_{k-1},x')p(t_{k-1},x'|Y(t_{k-1}))\left\{ d^nx' \right\}.
				\end{align}
				Note that $p(t_0|Y(t_0))$ is simply the initial condition $p(t_0,x_0)=\sigma_0(x_0)$.
			\item The measurement at time $t_k$ are incorporated in the correction step via
				\begin{align}
					p(t_k,x|Y(t_k))=\frac{p(y(t_k)|x)p(t_k,x|Y(t_{k-1}))}{\int p(y(t_k)|\xi)p(t_k,\xi|Y(t_{k-1}))}\left\{ d^n\xi \right\}.
				\end{align}
		\end{enumerate}
\end{enumerate}

\subsection{Practical Computational Strategies}\label{ssec:PracCompStrat}

The above general filtering algorithm based on the Dirac-Feynman approximation of the path integral formula computes the conditional probability density at grid points. This can be computationally very expensive as the number of grid points can be very large, especially for larger dimensions. Here, a few approximations will be presented that drastically reduces the computational load.   

The most crucial property that is exploited is that the transition probability density is an exponential function. Consequently,  many elements of the transitional probabilty tensor are negligible small, the precise number depending on the grid spacing. A significant computational saving is obtained when the (user-defined) ``exponentially small'' quantities are simply set to zero. In that case, the transition probability density is approximated by a sparse tensor, which results in huge savings in memory and computational load.

The next key issue is that of grid spacing. An appropriate grid spacing is one that adequately samples the conditional probability density. Of course, the conditional probability density is not known, but its effective domain (i.e., where it is significant)  is clearly a function of the signal and measurement model drifts and noises. For instance, the grid spacing should be of the order of change in state expected in a time step, which is not always easy to determine for a generic model. However, if the measurement noise is small, finer grid spacing is required so as to capture the state information in precise measurements. However, if the measurement noise is large, it may be unnecessary to use a fine grid spacing even if the state model noise is very small since the measurements are not that informative. Alternatively, If the grid spacing is too large compared to the signal model noise vielbein term, replace the difusion matrix with an ``effective diffusion matrix'' that is taken to be a constant times the grid spacing, i.e., noise inflation. This additional approximation can still lead to useful results as shown in an example below. 

It is also noted that the grid spacing is a function of the time steps. This is analogous to the case of PDE solution via discretization.  Thus, when using the one-step DF approximation, there will not be a gain by reducing the grid spacing to smaller  values (and at the cost of drastically increasing processing time). It is then more appropriate to use multi-step approximations to get more accurate results. 

Here are some possibilities for practical implemenetation:
\begin{enumerate}
	\item Precompute the transition probability tensor wth pre-determined and fixed grid. There are two options:
\begin{enumerate}
	\item Compute the correction at all the grid points;
	\item Compute the correction only where the prediction result is significant.
\end{enumerate}
It is the second of those options that will be used in this paper. 
\item Another option is to use a focussed adaptive grid, much as in PDE approaches. Specifically, at each time step:
\begin{enumerate}
	\item Find where the prediction step result is significant;
	\item Find the domain in the state space where the conditional probability density is significant, and possibly interpolate. For the multi-modal case, there would be several disjoint regions;
	\item Compute the transition probability tensor with those points as initial points and propagate to points in region suggested by state model.
\end{enumerate}
Thus, the grid is moving. In this case, the grid can be finer than in the previous case, although then the computational advantage of pre-computation is lost. 
\item Precompute the solution using basis functions. For instance, in many applications wavelets have been known to represent a wide variety of functions arising in practice. Then, instead of using the transition probability tensor, FPKfe solutions with wavelet basis functions are stored.
\end{enumerate}

\section{Examples}

In this section, a couple of two-dimensional examples are presented that illustrate the utility of the path integral formulae presented in this paper. The signal and measurement models are both nonlinear in these examples. Therefore, the Kalman filter is not a reliable solution for these problems. The symmetric discretization formula was used. The MATLAB tensor toolbox developed by Bader and Kolda was used for the computations \cite{ACM-TOMS-TENSORTOOLBOX}. The approximation techniques are discussed in Section \ref{ssec:PracCompStrat}. In addition, in order to speed up the pre-computation of the transition probability tensor, it was assumed that $P(f_1,f_2|i_1,i_2)=0$ if $|f_r-i_r|>2$, i.e., the ``extent'' of $P$ was chosen to be 2. Thus, this implementation of the DF algorithm is sub-optimal in many ways.

 For comparison, the performance of the SIR particle filter based on the Euler discretization of the state model SDE is also included\cite{BudhirajaaChenLee2007}. The MATLAB toolbox PFlib was used in the particle filter simulations\cite{LingjiChenandChihoonLeeandAmarjitBidhirajaandRamanK.Mehra2007}. 
\subsection{Example 1}\label{ssec:FungRoz}

\FIGURE{\centering
		\scalebox{0.75}{\includegraphics{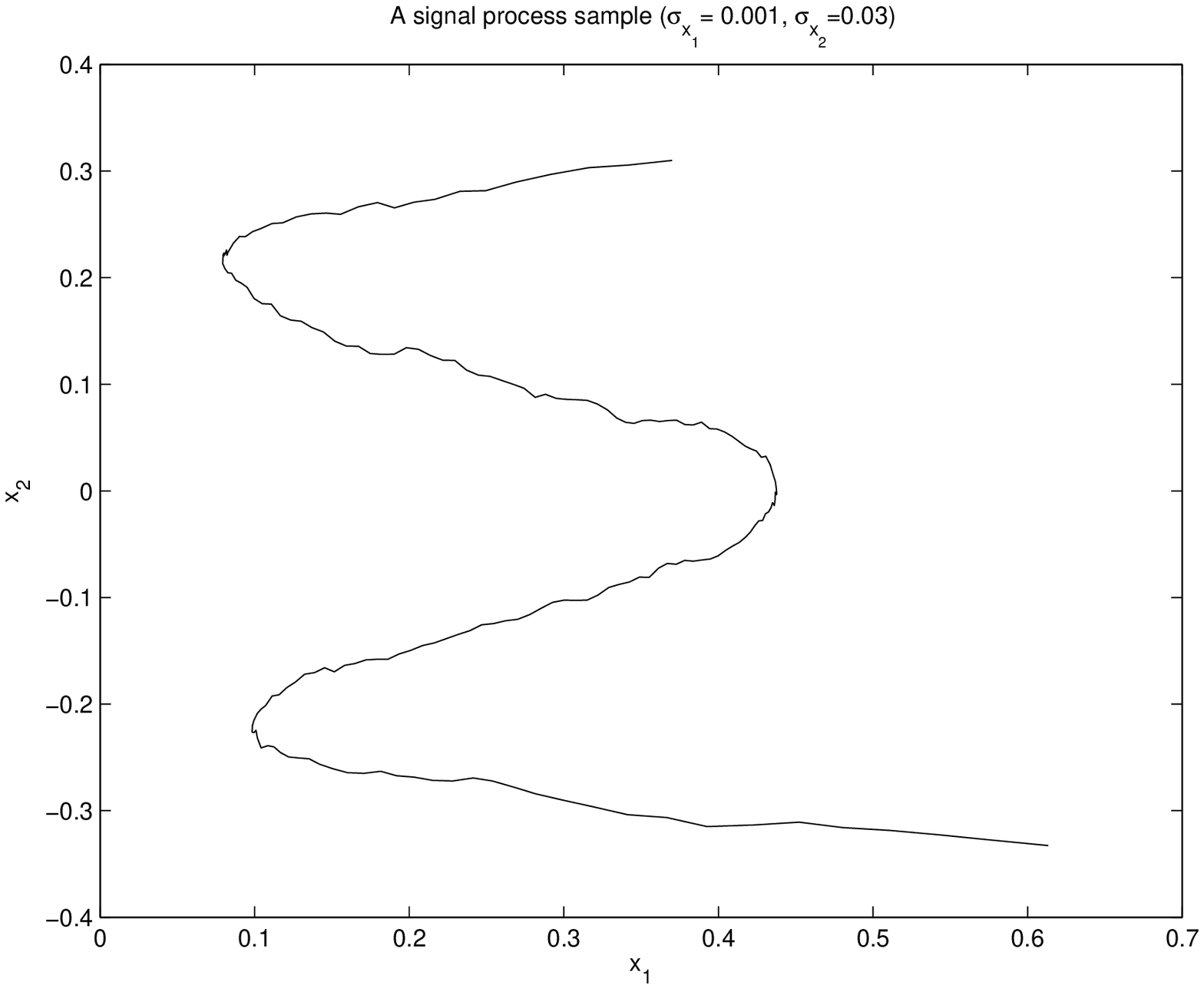}}
	\caption{A Sample trajectory for the state model in Equation \ref{eq:StateModelExample1}.}
	\label{fig:FungRozCDSigSamp}
}

\FIGURE{\centering
		\scalebox{0.75}{\includegraphics{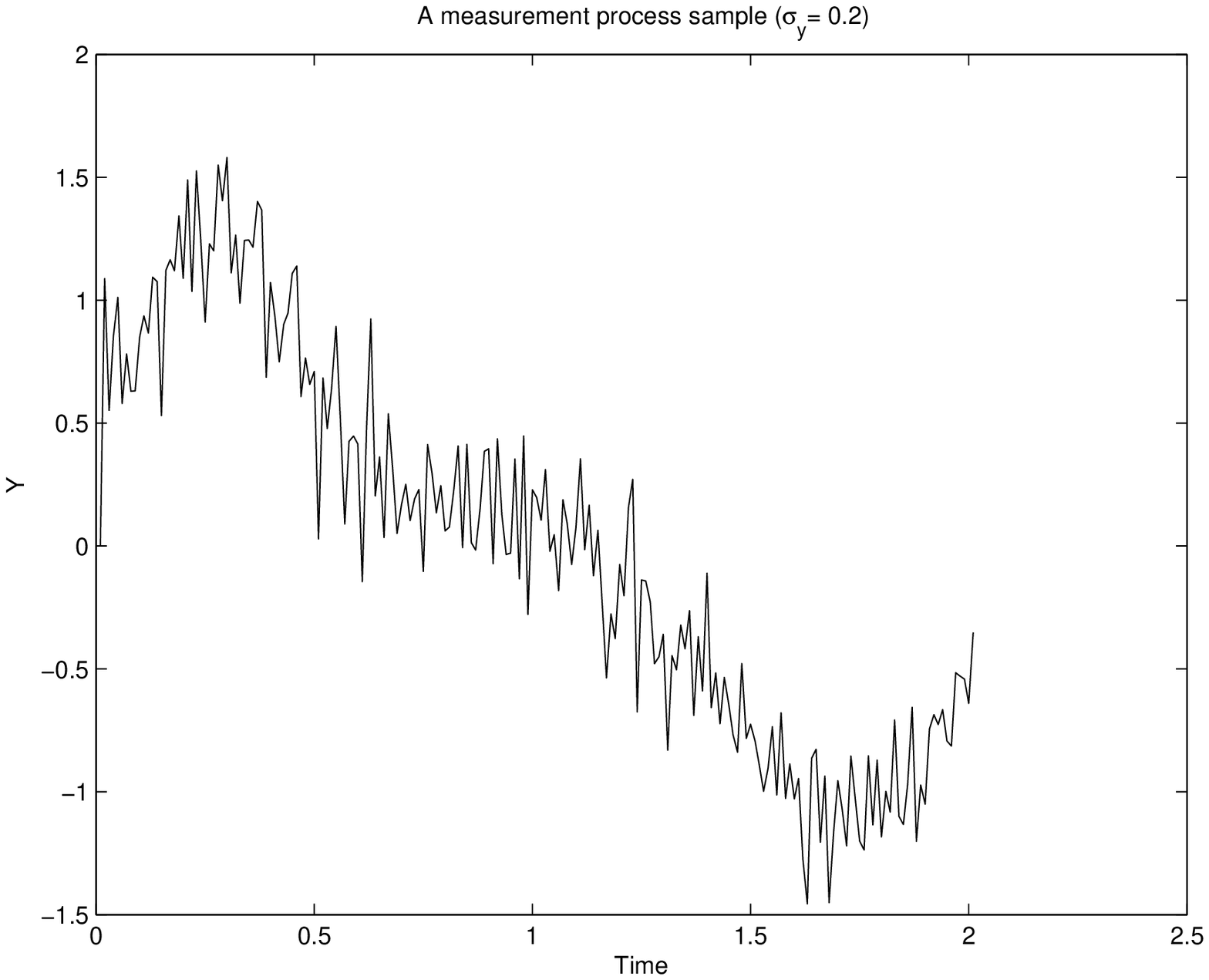}}
	\caption{A measurement sample for the state trajectory in Figure \ref{fig:FungRozCDSigSamp} and measurement model given by Equation \ref{eq:MeasModelExample1}.}
	\label{fig:FungRozCDMeasSamp}
}

Consider the state model
\begin{align}\label{eq:StateModelExample1}
	d\rv{x}_1(t)&=\left( -189\rv{x}_2^3(t)+9.16\rv{x}_2(t) \right)dt+\sigma_{\rv{x}_1}d\rv{v}_1(t),\\ \nonumber
	d\rv{x}_2(t)&=-\frac{1}{3}dt+\sigma_{\rv{x}_2}d\rv{v}_2(t),
\end{align}
with the nonlinear measurement model
\begin{align}\label{eq:MeasModelExample1}
	\rv{y}(t_k)=\sin^{-1}\left( \frac{\rv{x}_2(t_k)}{\sqrt{\rv{x}_1^2(t_k)+\rv{x}_2(t_k)}} \right)+\sigma_{\rv{y}}\rv{w}(t_k).
\end{align}
Here $\begin{bmatrix}\sigma_{\rv{x}_1}&\sigma_{\rv{x}_2}\end{bmatrix}=\begin{bmatrix}0.001&0.03\end{bmatrix}$ and we consider two values for $\sigma_{\rv{y}}$, namely $\sigma_{\rv{y}}=0.2$ and $\sigma_{\rv{y}}=2$. This example was studied in \cite{S.V.LototskyB.L.Rozovskii1998} and the extended Kalman is known to fail for this model.

\FIGURE{\centering
		\scalebox{0.75}{\includegraphics{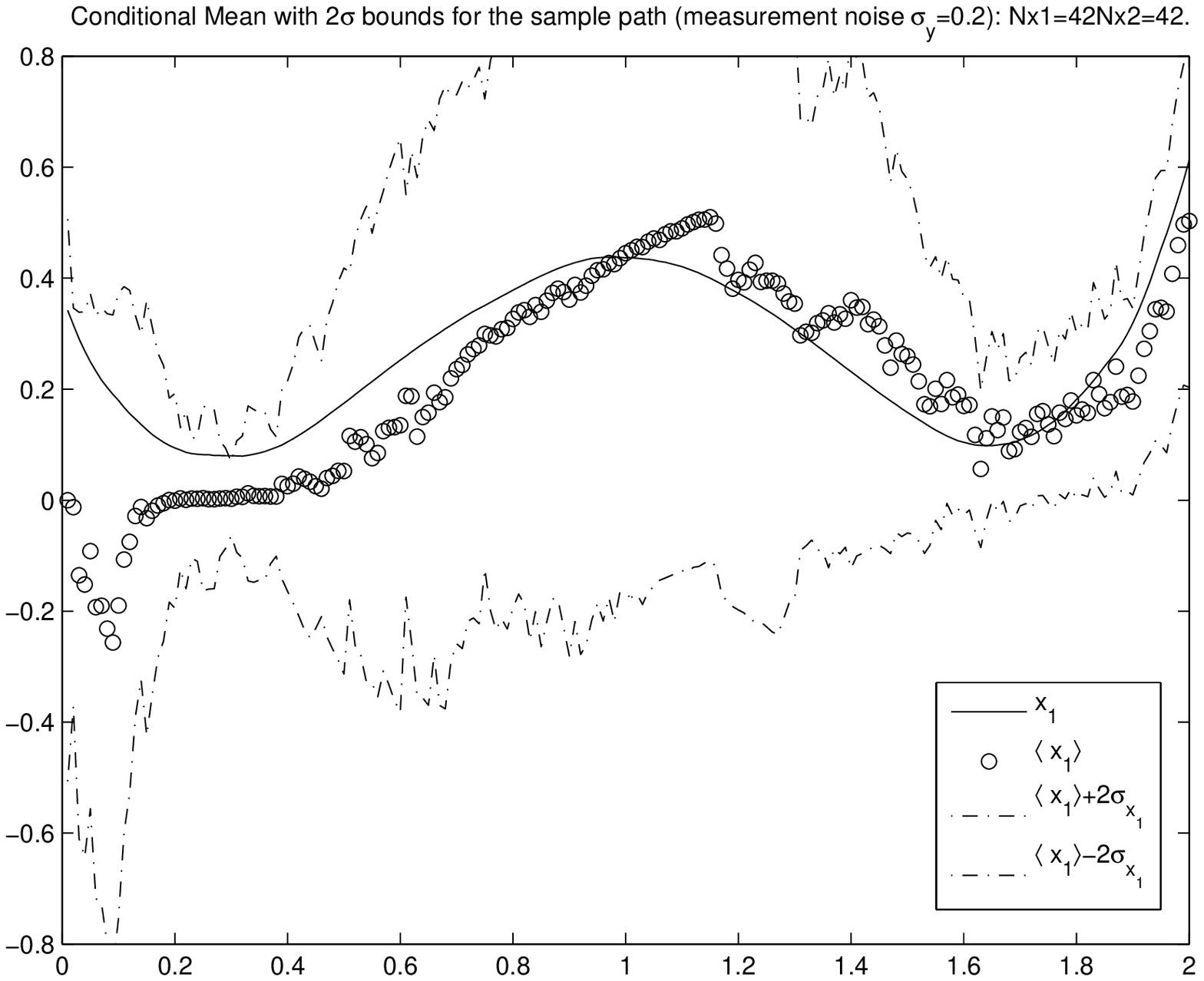}}
	\caption{Conditional mean $\rv{x}_1(t)\rangle$ computed for the measurement sample of Figure \ref{fig:FungRozCDSigSamp}.}
	\label{fig:FungRozCDCMx1}
}

\FIGURE{\centering
		\scalebox{0.75}{\includegraphics{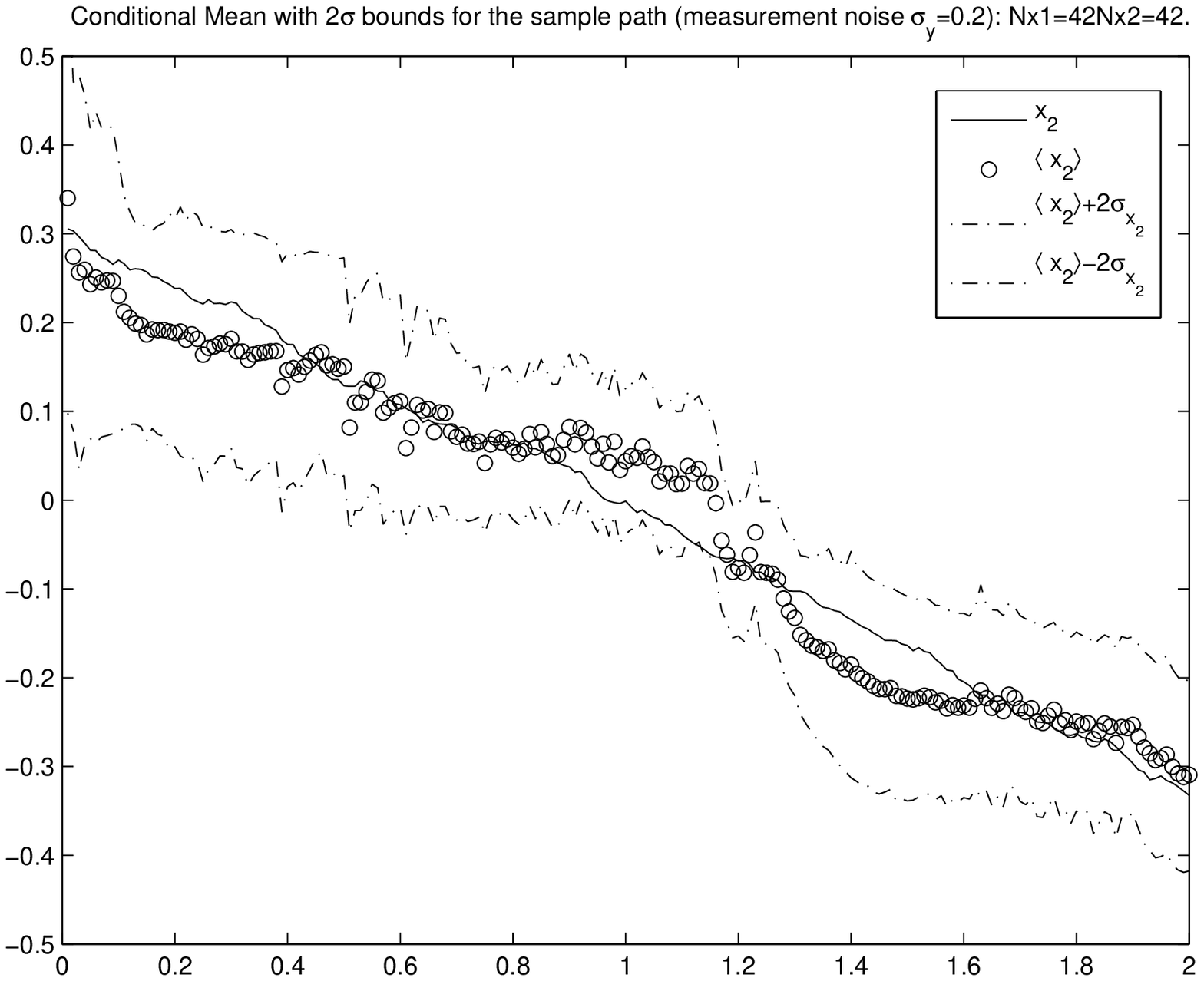}}
	\caption{Conditional mean $\rv{x}_2(t)\rangle$ computed for the measurement sample of Figure \ref{fig:FungRozCDSigSamp}.}
	\label{fig:FungRozCDCMx2}
}

\FIGURE{\centering
		\scalebox{0.75}{\includegraphics{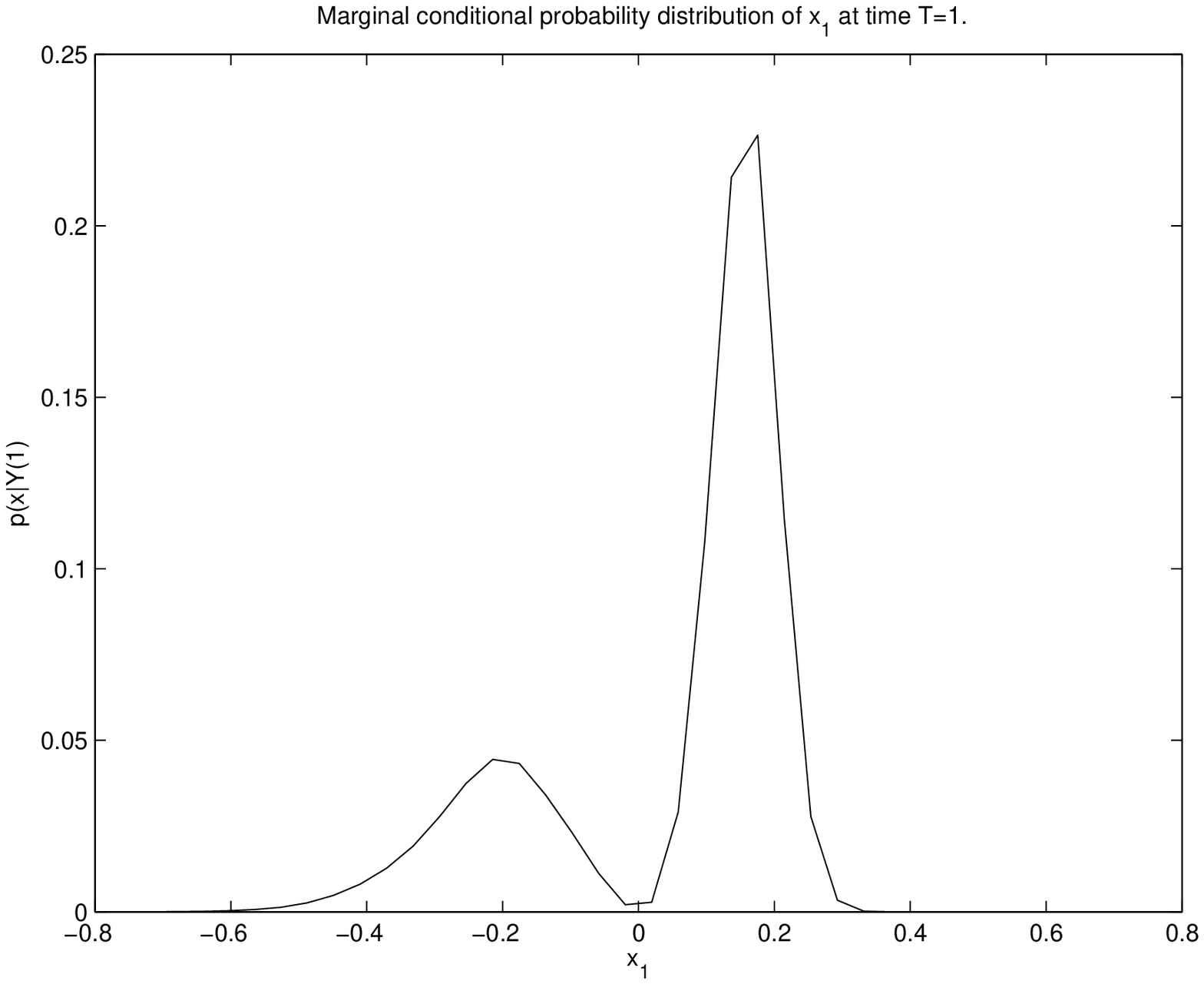}}
	\caption{Marginal conditional probability density for $\rv{x}_1(t)$.}
	\label{fig:FungRozBiModal}
}

The Lagrangian for this model is easily seen to be given by
\begin{align}
	\frac{1}{2\sigma_{x_1}^2}\left( \dot{x}_1(t)+189x_2^3(t)-9.16x_2(t) \right)^2+\frac{1}{2\sigma_{x_2}^2}\left( \dot{x}_2(t) +\frac{1}{3}\right)^2.
\end{align}

Consider first the $\sigma_{\rv{y}}=0.2$ case.  Figure \ref{fig:FungRozCDSigSamp} shows a sample trajectory with initial state $[0.37, 0.31]$, and in Figure \ref{fig:FungRozCDMeasSamp} is shown a measurement sample. The time step size is $0.01$ and the number of time steps is $200$. 

The spatial interval $[-0.8, 0.8] \times [-0.8, 0.8]$ is subdivided into $42\times42$ equal intervals. The signal model noise is very small requiring much finer grid spacing. Instead, as discussed in Section \ref{ssec:PracCompStrat}, the effective $\sigma$'s were taken to be $\alpha\times\begin{bmatrix}\Delta x_1&\Delta x_2\end{bmatrix}$ with $\alpha= 1$.  The initial distribution is taken to be uniform. 

	The conditional mean for $\rv{x}_1(t)$ and $\rv{x}_2(t)$ are plotted in Figures \ref{fig:FungRozCDCMx1} and \ref{fig:FungRozCDCMx2}. The RMS error was found to be  0.1180 and the time taken was 40 seconds. Observe that the conditional mean is within a standard deviation of the true state almost all of the time confirming that the tracking quality is good. It is noted that the variance is larger for $\rv{x}_1(t)$. The reason can be understood from Figure \ref{fig:FungRozBiModal} which plots the marginal conditional probability density of the state variable $\rv{x}_1(t)$--- it is bi-modal. The bi-modal nature (for a significant fraction of the time) is the reason the EKF will fail in this instance. The performance is seen to be similar to that reported in \cite{S.V.LototskyB.L.Rozovskii1998} which was obtained using considerably more involved techniques and finer grid spacing.  

	This example was also investigated using the SIR-PF. The SIR-PF implemented with 5000 particles took about 155 seconds and the RMS error was found to be $0.165$ even when initiated about the true state, i.e., initial distribution was chosen to be  Gaussian with mean  $[0.37, 0.31]$ and variance $I_{2}$, where $I_2$ is the $2\times2$ identity matrix. When the variance was reduced to $10^{-2}\times I$, it resulted in RMS error of only $0.022$. Thus, the performance of the SIR-PF depends crucially on the initial condition. It is also noted that no bi-modality of the marginal pdf of state $\rv{x}_1(t)$ at $T=1$ was observed for the SIR-PF simulations when the number of particles was 5000. Upon increasing the number of particles to $10,000$, the bi-modality was noted, although the RMSE was not significantly smaller.

\FIGURE{\centering
		\scalebox{0.75}{\includegraphics{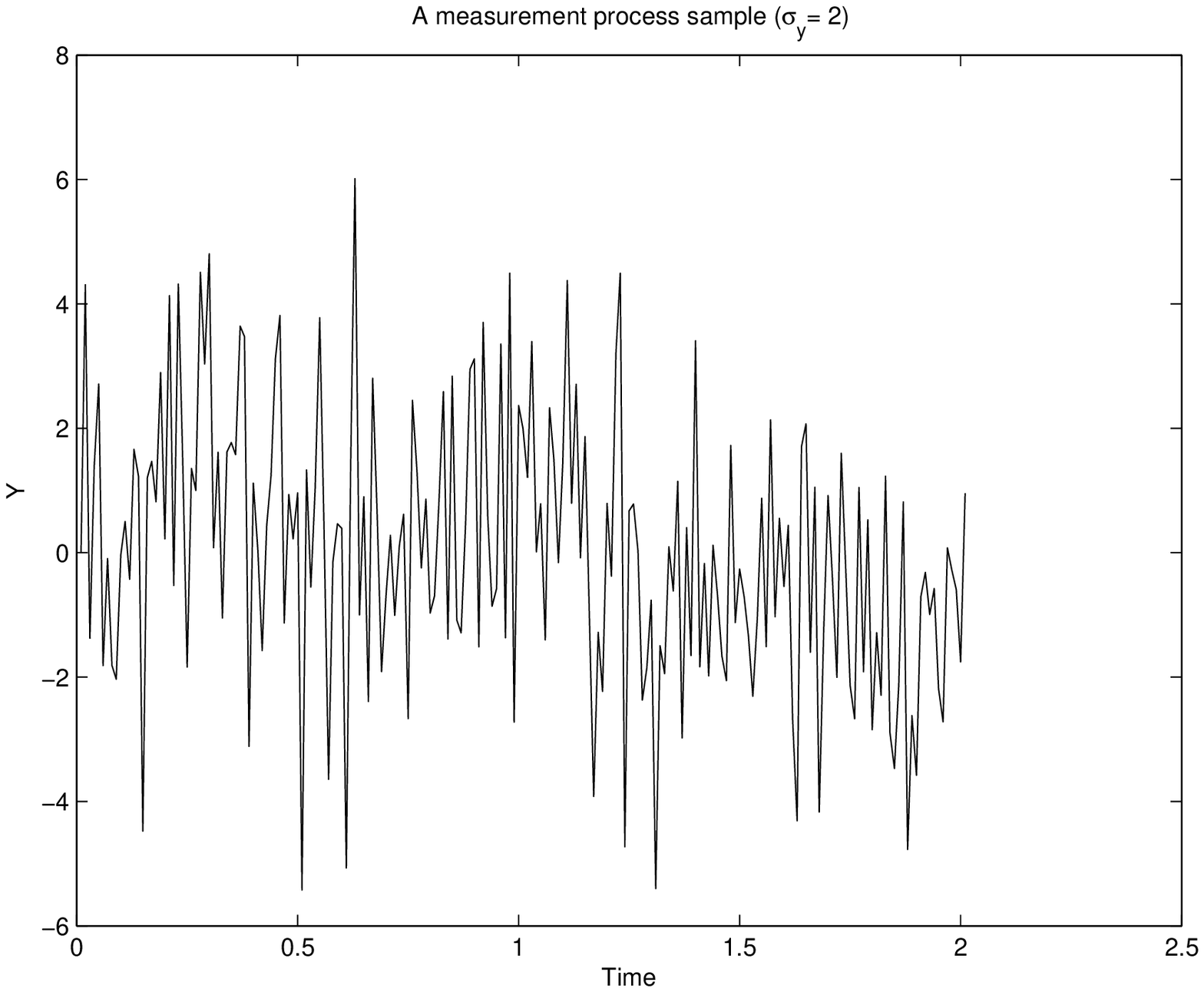}} 
	\caption{A measurement sample for the state trajectory in Figure \ref{fig:FungRozCDSigSamp} for larger measurement noise ($\sigma_{\rv{y}}=2$.)}
	\label{fig:EX1MeasSampsigy2}
}

\FIGURE{\centering
	\scalebox{0.75}{\includegraphics{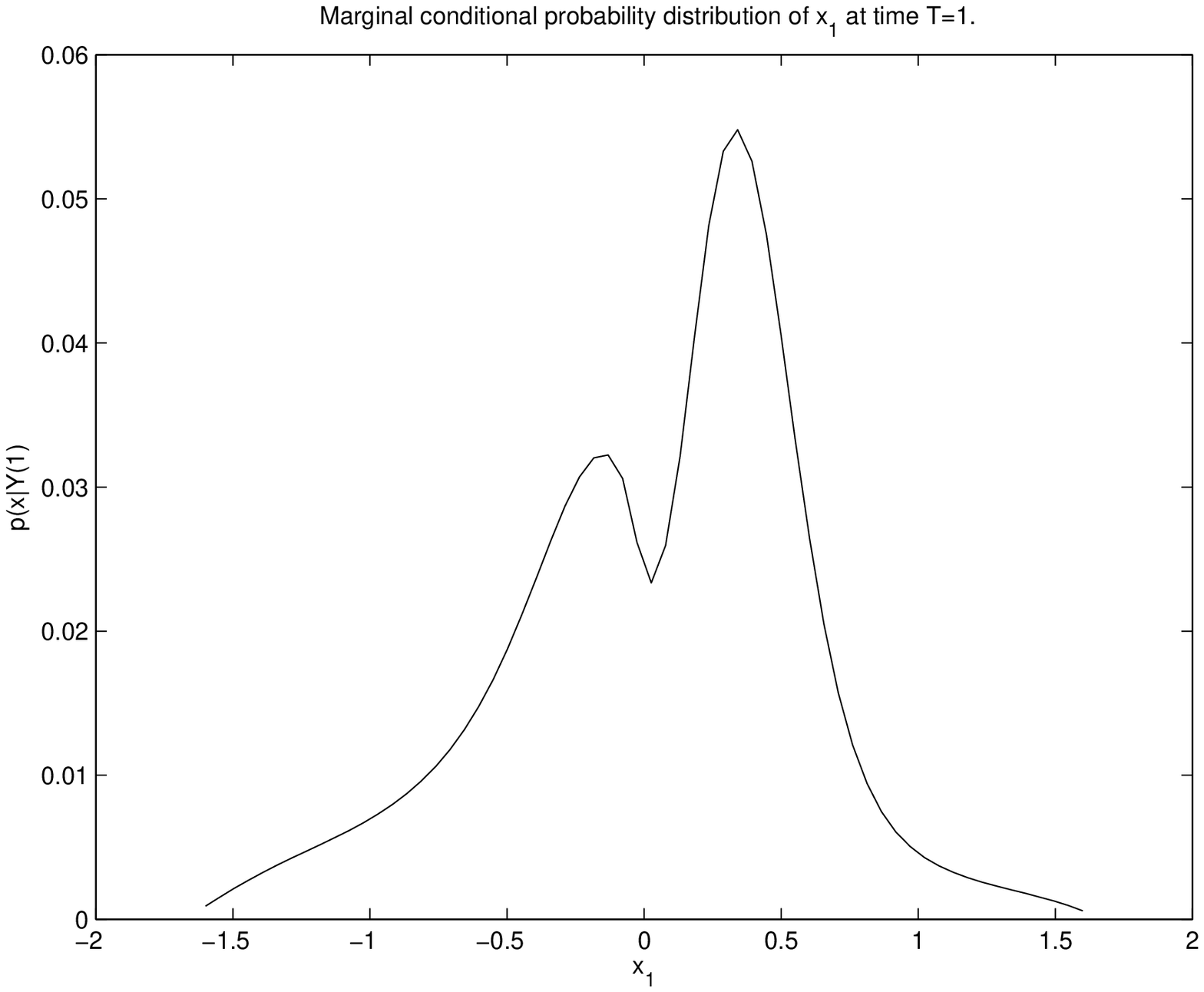}}
	\caption{Marginal conditional probability density for $\rv{x}_1(t)$ for the larger measurement noise case..}
	\label{fig:Ex1BiModalsigy2}
}

Next consider the larger measurement noise case. A measurement sample corresponding to the state trajectory in Figure \ref{fig:FungRozCDSigSamp} is shown in Figure \ref{fig:EX1MeasSampsigy2}. The spatial interval $[-1.6, 1.6] \times [-1, 1]$ was subdivided into $62\times62$ equispaced grid points. The RMS error was found to be 0.128 and the time taken was about 110 seconds. The bimodality at $T=1$ is evident in this case as well (see Figure \ref{fig:Ex1BiModalsigy2}).  

The SIR-PF was also implemented. When initialized as Gaussian with zero mean and unit variance, the tracking performance of the SIR-PF failed; the RMSE was found to be 25.34 when using 5000 particles (taking about 110 seconds).  A sample performance is shown in Figures \ref{fig:EX1SIRPFCMx1sigy2} and \ref{fig:EX1SIRPFCMx2sigy2}; it is clear that the state $\rv{x}_1$ is poorly tracked. Even when using  50,000 particles, a sample run that took 25 minutes, resulted in RMS error of 16.53.

\FIGURE{\centering
		\scalebox{0.75}{\includegraphics{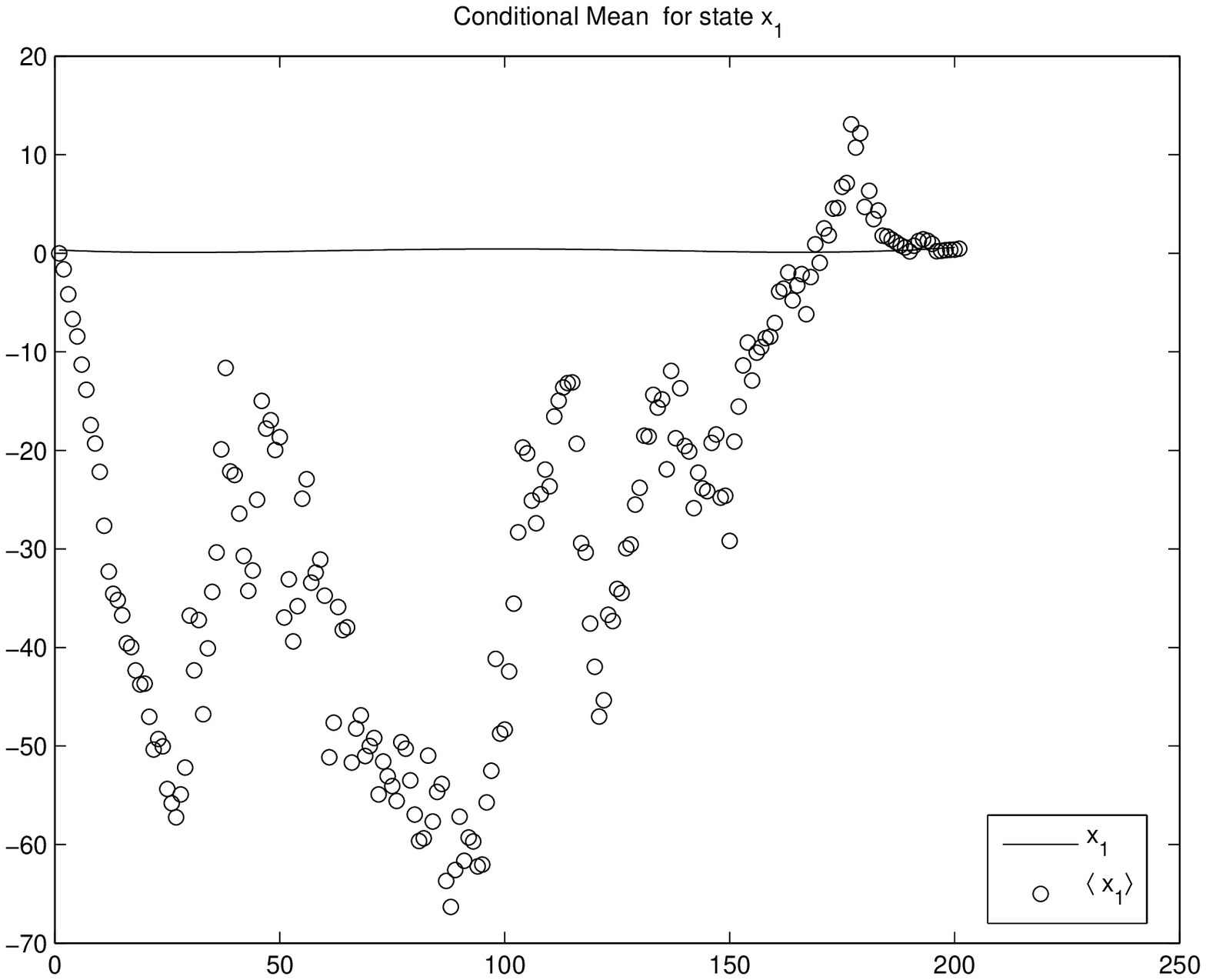}}
	\caption{Conditional mean for state $\rv{x}_1(t)$ computed using 5000 particles for $\sigma_{\rv{y}}=2$.}
	\label{fig:EX1SIRPFCMx1sigy2}
}

\FIGURE{\centering
		\scalebox{0.75}{\includegraphics{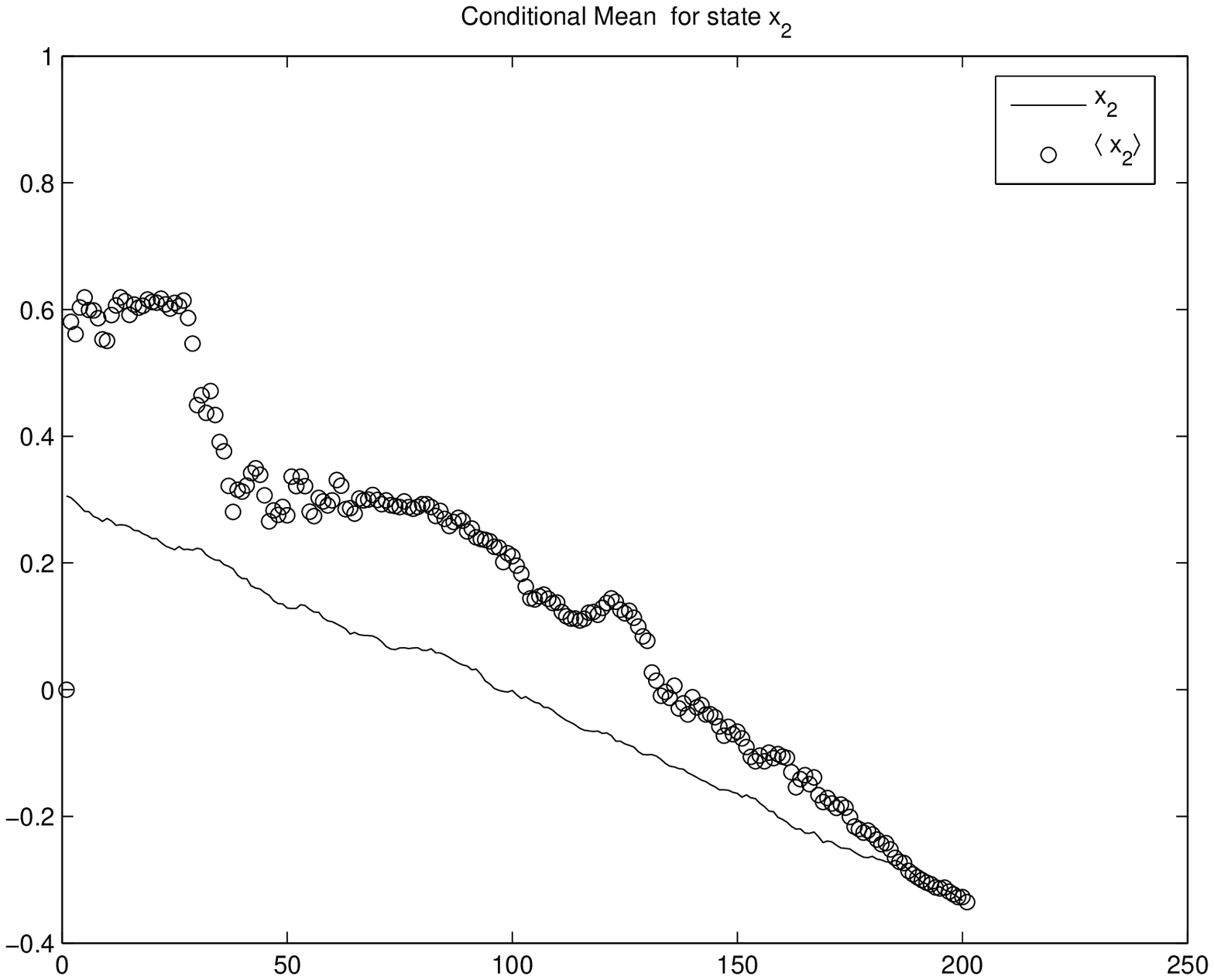}}
	\caption{Conditional mean for state $\rv{x}_2(t)$ computed using 5000 particles for $\sigma_{\rv{y}}=2$.}
	\label{fig:EX1SIRPFCMx2sigy2}
}

\clearpage
\subsection{Example 2}

\FIGURE{\centering
	\includegraphics{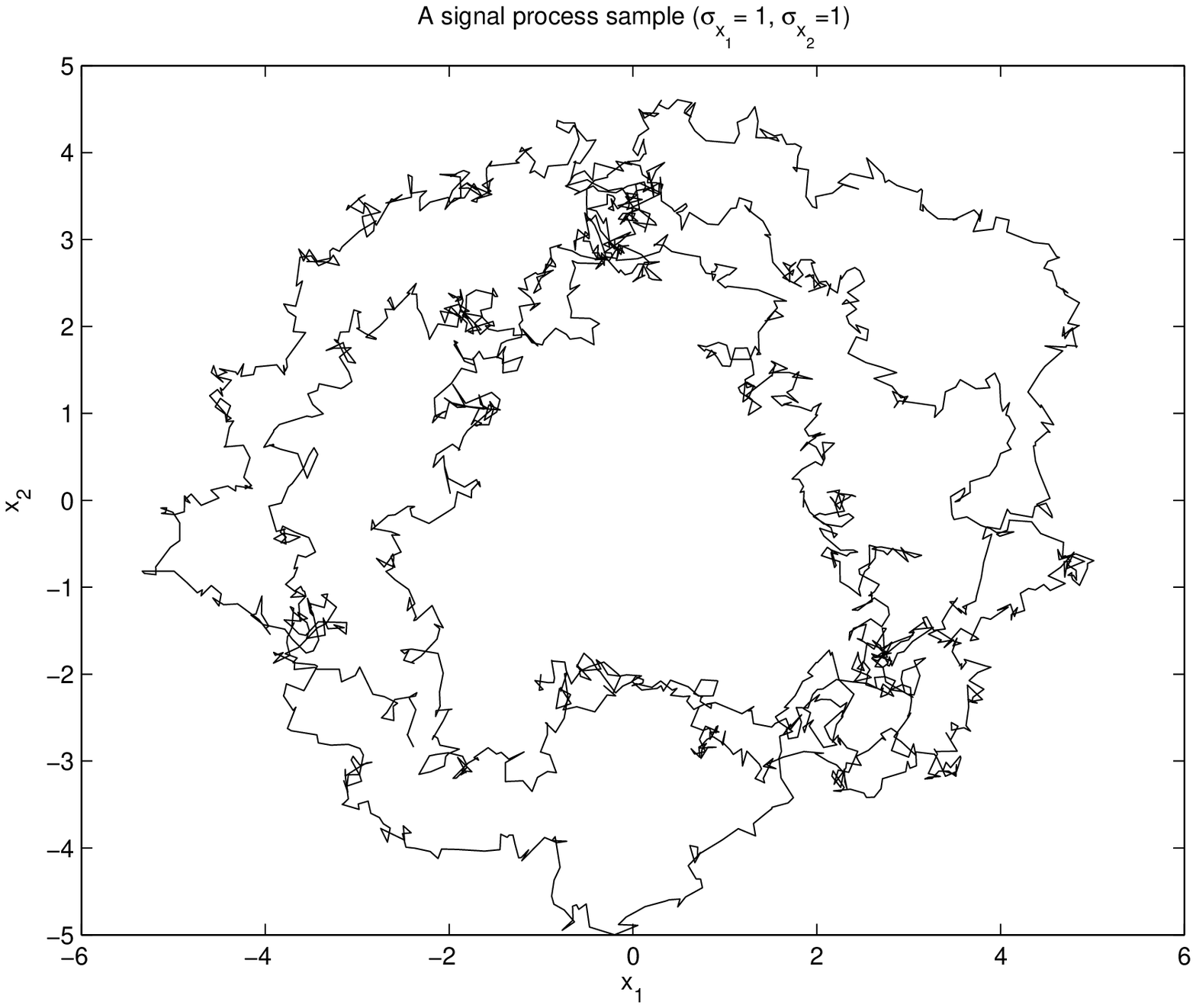}
	\caption{A sample trajectory for the state model in Equation \ref{eq:StateModelExample2}.}
	\label{fig:EX2StateModel}
}

\FIGURE{\centering		
	\includegraphics{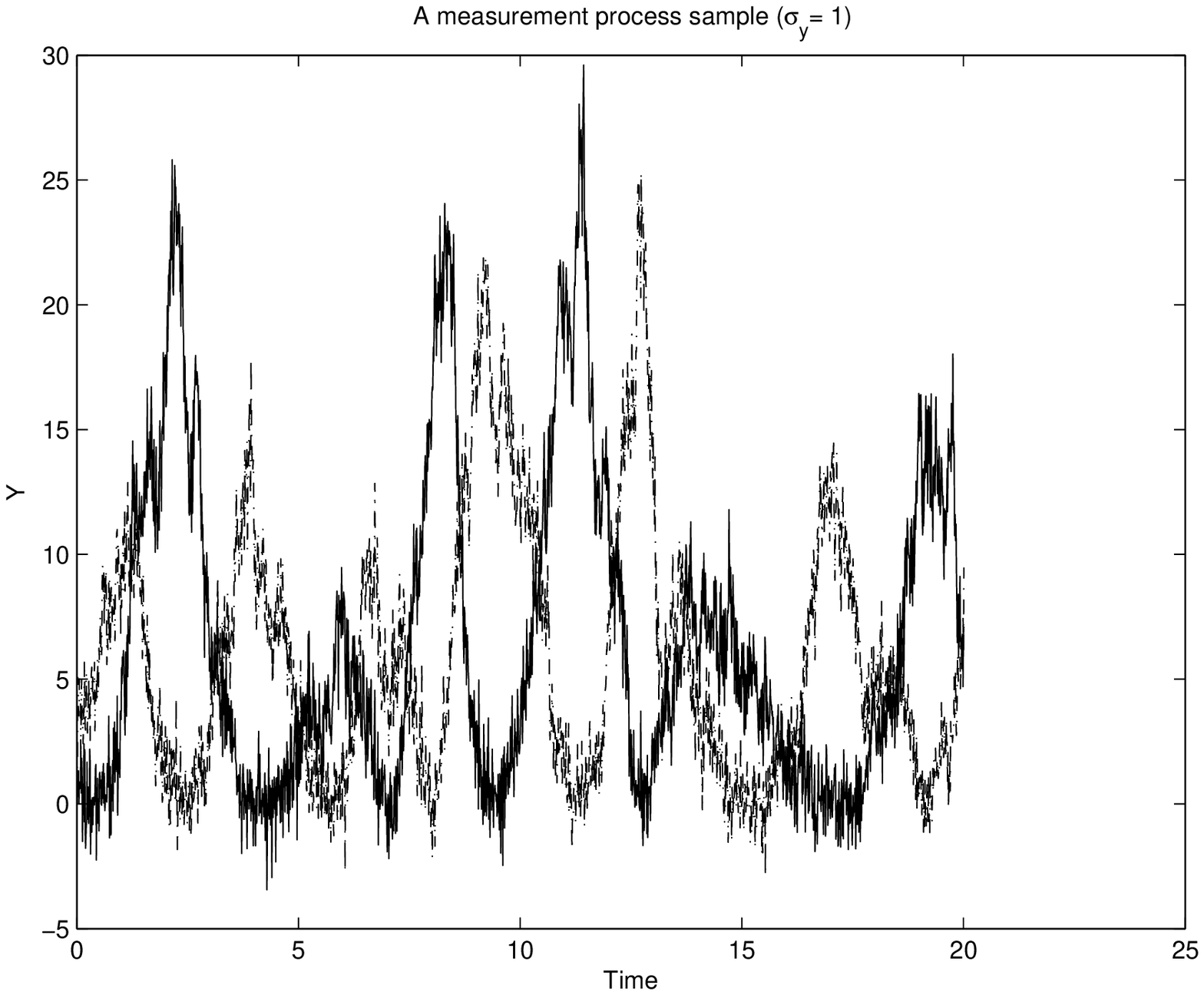}
	\caption{A measurement sample for the state trajectory in Figure \ref{fig:EX2StateModel} and measurement model given by Equation \ref{eq:MeasurementModelExample2}.}
	\label{fig:EX2MeasSample}
}

Consider the  state model
\begin{align}
\label{eq:StateModelExample2}
	d\rv{x}_1(t)&=(-\rv{x}_2(t)+\cos(\rv{x}_1(t))dt+d\rv{v}_1(t),\\ \nonumber
	d\rv{x}_2(t)&=(\rv{x}_1(t)+\sin(\rv{x}_2(t)))dt+d\rv{v}_2(t),  \nonumber
\end{align}
and the measurement model
\begin{align}
\label{eq:MeasurementModelExample2}
	d\rv{y}_1(t_k)&=\rv{x}_1^2(t_k)+\rv{w}_1(t_k),\\ \nonumber
	d\rv{y}_2(t_k)&=\rv{x}_2^2(t_k)+\rv{w}_2(t_k).  \nonumber
\end{align}
Here $d\rv{v_i}(t)$ are uncorrelated standard Wiener processes, and $w_i(t_k)\sim N(0,1)$. A sample of the state and measurement processes are shown in Figures \ref{fig:EX2StateModel} and \ref{fig:EX2MeasSample} respectively. The discretization time step is 0.01.  

The initial distribution is taken to be a Gaussian with zero mean and variance of 10. Figures \ref{fig:EX2CMx1-0-10} and \ref{fig:EX2CMx2-0-10} plots the conditional mean for the state. It is seen that the tracking is quite good despite the error at the start; the RMS error was found to be 0.54. The interval $[-6,6]$ was uniformly divided into 62 grid points and the extent of the transition probability tensor was 2.  The 2000 time steps took about 8 minutes. 

The SIR particle filter was also implemented with the same initial condition and with $5000$ particles. The RMS error was found to be $1.48$.  Each run took about 10 minutes. 

\FIGURE{\centering		
	\scalebox{0.25}{\includegraphics{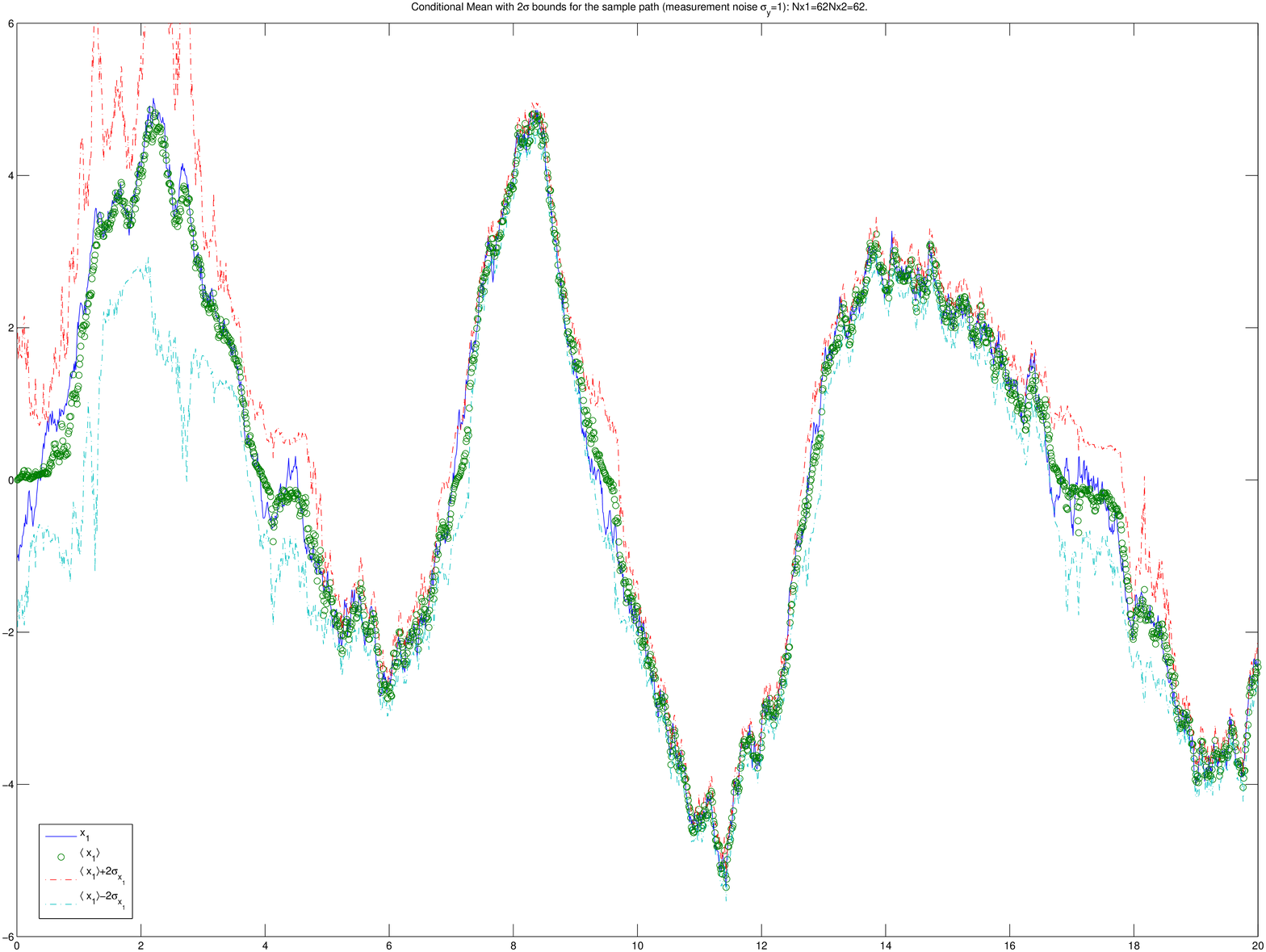}}
	\caption{Conditional mean for state $\langle\rv{x}_1(t)$ computed for the measurement sample of Figure \ref{fig:EX2MeasSample} and with initial distribution $N(0,10$.}
	\label{fig:EX2CMx1-0-10}
}

\FIGURE{\centering
	\scalebox{0.25}{\includegraphics{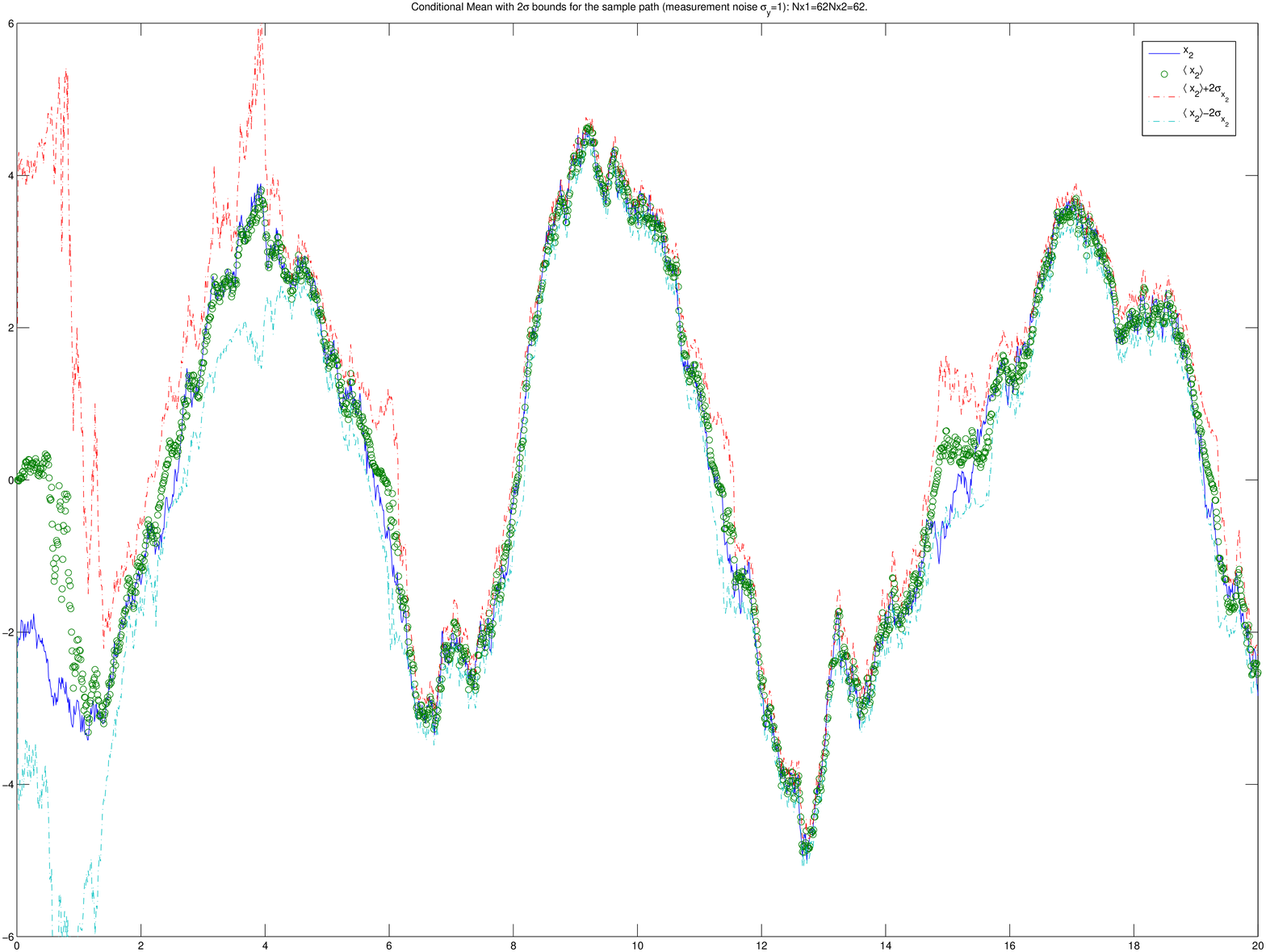}}
	\caption{Conditional mean for state $\langle\rv{x}_2(t)$ computed for the measurement sample of Figure \ref{fig:EX2MeasSample} and with initial distribution $N(0,10$.}
	\label{fig:EX2CMx2-0-10}
}

Next, consider time step of $0.2$, i.e., only every twentieth measurement sample is assumed given. Figures  \ref{fig:EX2CMx1p20} and \ref{fig:EX2CMx2p20} show the conditional means of the states. The number of grid points is smaller; the grid spacing is chosen to be twice the previous instance. Consequently, the computational effort is less, requiring only about 14 seconds. It is noted that the tracking performance is very good and  the error estimated form the conditional probability density using this approximation is reliable. Now the RMS error is found to be 0.69, and only 0.31 if the first few errors are ignored. 

\FIGURE{\centering
		\includegraphics{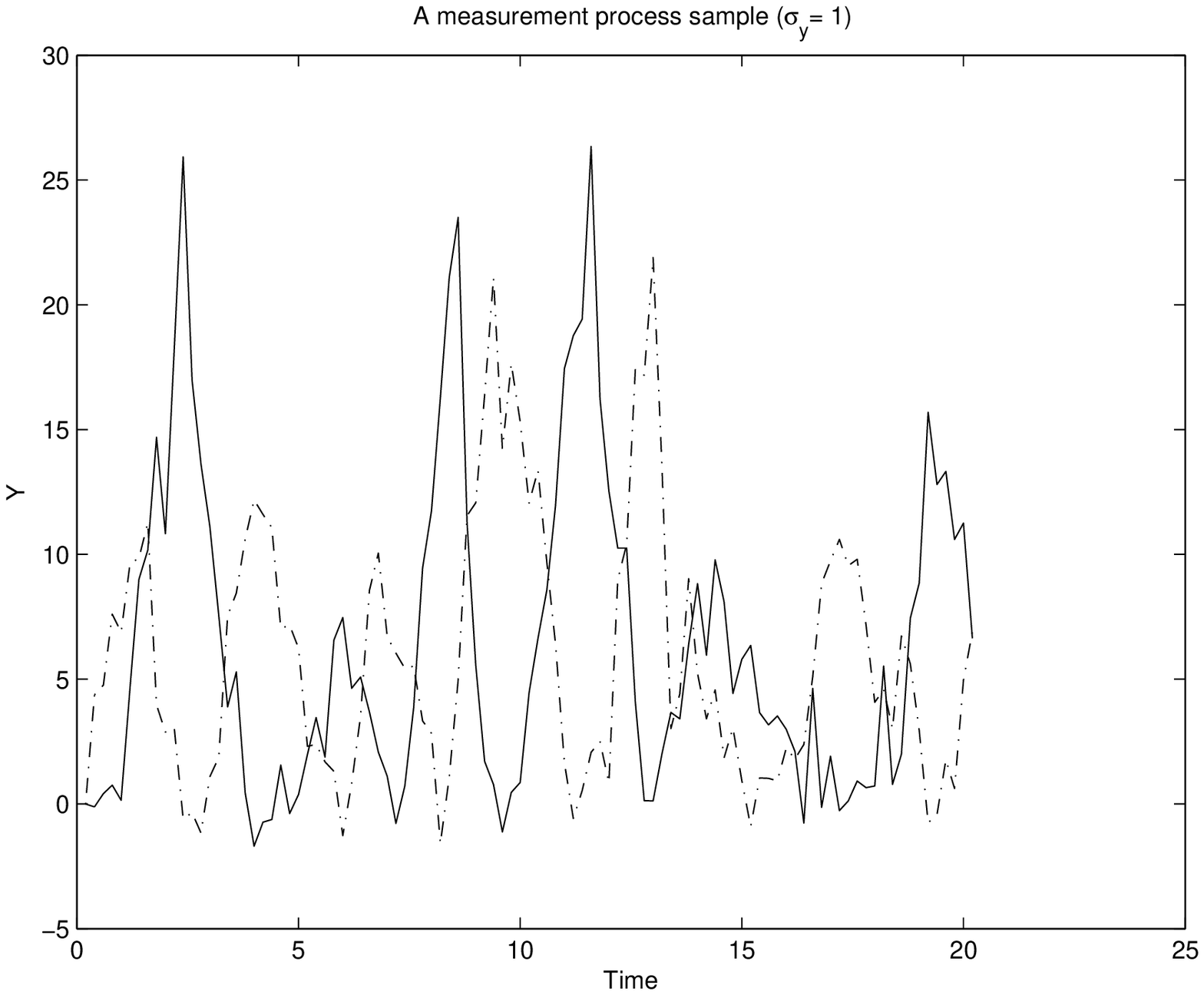}
	\caption{A measurement sample for the state trajectory in Figure \ref{fig:EX2StateModel} and measurement model given by Equation \ref{eq:MeasurementModelExample2} with measurement interval of $T=0.20$.}
	\label{fig:EX2MeasSamplep20}
}

\FIGURE{\centering		
	\scalebox{0.75}{\includegraphics{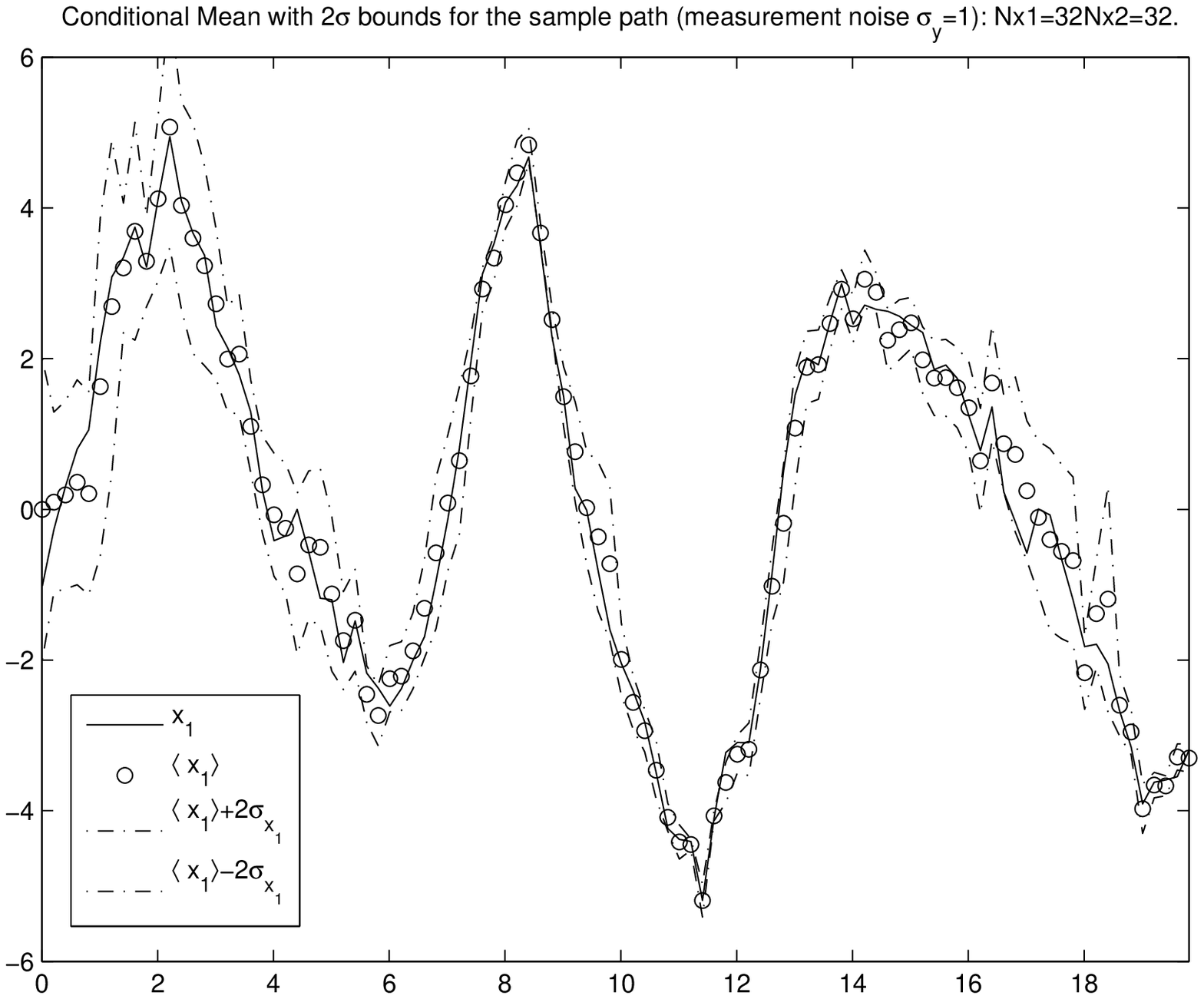}}
	\caption{Conditional mean for state $\langle\rv{x}_1(t)$ when measurement sample given by Figure \ref{fig:EX2MeasSamplep20}. }
	\label{fig:EX2CMx1p20}
}

\FIGURE{\centering
		\scalebox{0.75}{\includegraphics{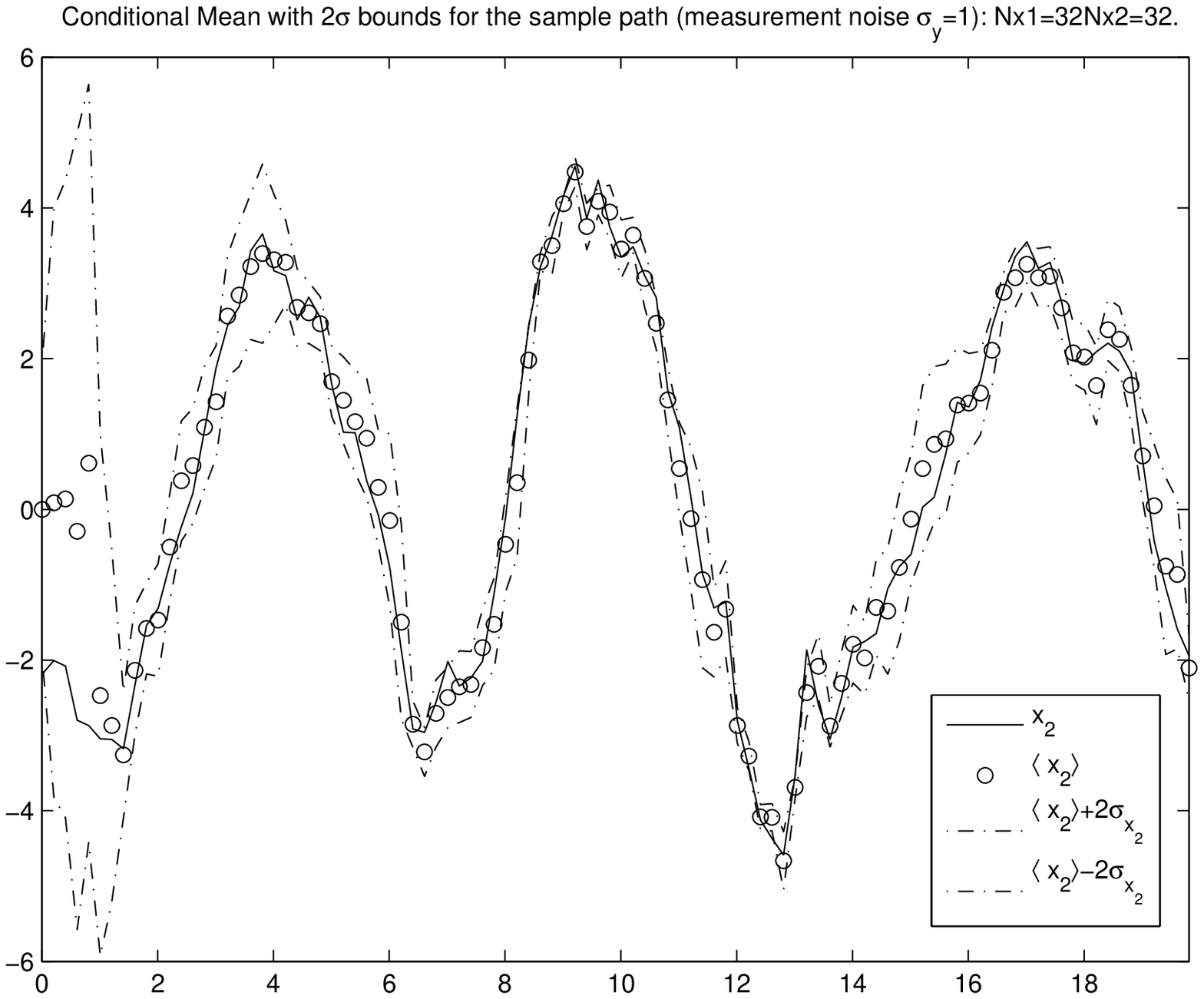}}
	\caption{Conditional mean for state $\langle\rv{x}_2(t)$ when measurements are every $0.2$  seconds (Figure \ref{fig:EX2MeasSamplep20}). }
	\label{fig:EX2CMx2p20}
}

In contrast, the error using the SIR-PF (with 2000 particles that took 37 seconds) is found to fail with RMS error of $3.68$; see Figures \ref{fig:EX2SIRPFCMx1p20} and \ref{fig:EX2SIRPFCMx2p20} for typical results. The results for increasing the number of particles to $50,000$ (14 minutes execution time) did not improve the situation significantly (RMS error of $3.34$); it would be better to subdivide the time step into several time steps and do the SIR-PF. For the purposes of this paper, it is sufficient to note that in this instance a single-step DF algorithm succeeds where the one-step SIR-PF fails. 
\FIGURE{\centering
	\scalebox{0.75}{\includegraphics{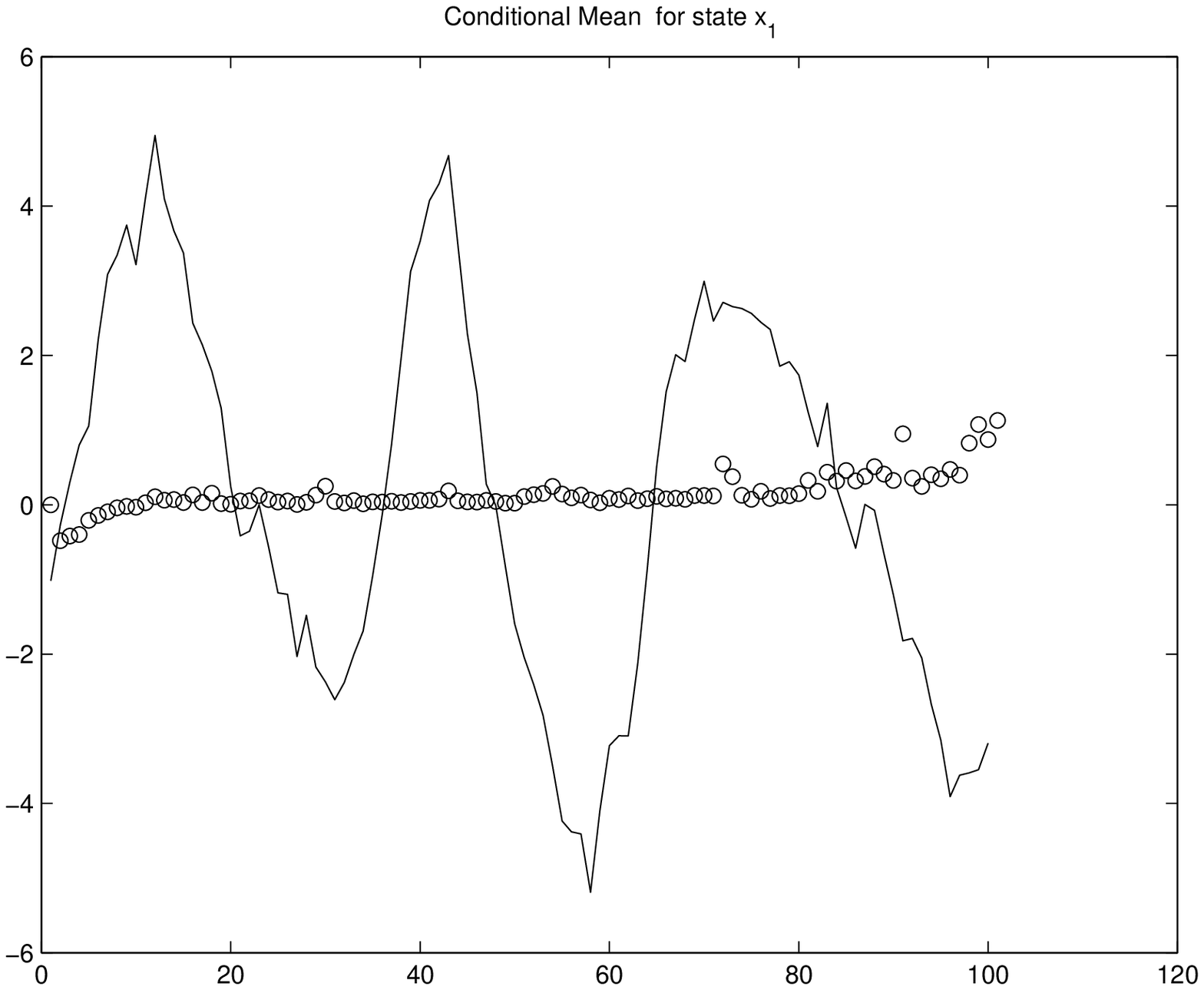}}
	\caption{Conditional mean for state $\langle\rv{x}_1(t)$ computed using the SIR-PF when measurement sample given by Figure \ref{fig:EX2MeasSamplep20}. }
	\label{fig:EX2SIRPFCMx1p20}
}

\FIGURE{\centering
		\scalebox{0.75}{\includegraphics{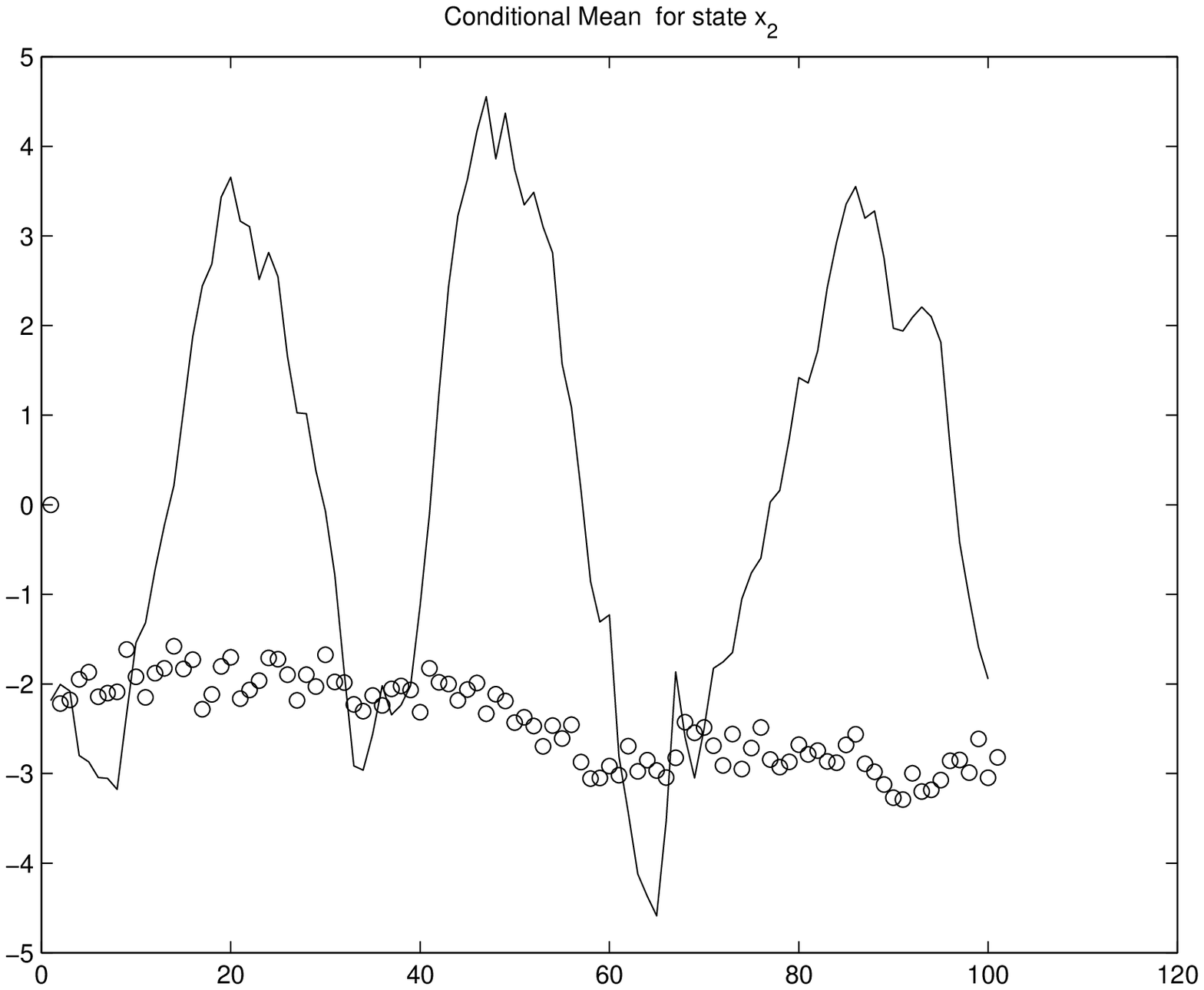}}
	\caption{Conditional mean for state $\langle\rv{x}_2(t)$ computed using the SIR-PF when measurements are every $0.2$  seconds (Figure \ref{fig:EX2MeasSamplep20}). }
	\label{fig:EX2SIRPFCMx2p20}
}

\clearpage
\section{Additional Remarks}\label{sec:AddlRemarks}

It is remarkable to note that the simplest approximations to the path integral formulae leads to very accurate results. Note that the time steps are small, but not infinitesimal. Such time step sizes are not unrealistic in real world applications.

It is particularly noteworthy since it was found that SIR-PF was not a reliable solution to the studied problems. Note that the rigorous results for MC type of techniques assume that the signal model drifts are bounded.  If that is not the case, as here, the SIR-PF is not guarenteed to work well. In any case, the speed of convergence to the correct solution is not specified for a general filtering problem, as emphasized in \cite{S.MaskellM.BriersR.WrightP.Horridge2005}; PFs need to be ``tuned'' to the problem to get desired level of performance. In fact, for discrete-time filtering problems, excellent performance also follows from a well-chosen grid using sparse tensor techniques \cite{BhashyamBalaji2008}. Clearly, it is not axiomatic that a generic particle filter will lead to significant computation savings (or performance) over a well-chosen sparse grid method, at least for smaller dimensional problems.  


Observe also that the DF path integral filtering formulae have a simple and clear physical interpretation. Specifically, when the signal model noise is small the transition probability is significant only near trajectories satisfying the noiseless equation. The noise variance quantifies the extent to which the state may deviate from the noiseless trajectory.

The following additional observations can be made on the simplest path integral filtering method proposed in this paper:
\begin{enumerate}
	\item In universal nonlinear filtering, including path integral filtering discussed here, the standard deviation computed from the (accurately computed) conditional probability density is a reliable measure of the filter performance. This is not the case for suboptimal methods like the EKF.
	\item In the examples studied, only the simplest one-step approximate formulae for the path integral expression were applied. There is a large body of work on more accurate one-step formulae that could be used to get better results if the formulae used in this paper are not accurate enough (see, for instance, \cite{F.LangoucheandD.RoekaertsandE.Tirapegui1982}) .
	\item  Observe that higher accuracy (than the DF approximation) is attained by approximating the path integral with a finite-dimensional integral. The most efficient technique for evaluating such integrals would be to use Monte Carlo or quasi Monte Carlo methods. Another possibility is to use Monte Carlo based techniques for computing path integrals \cite{G.PeterLepage2000}. Observe that this is different from particle filtering. 
	\item The major source of computational savings following from noting that the transition probability is given in terms of an exponential function. This implies that $P(t'', x''|t', x')$ is non-negligible only in a very small region, or the transition probability density tensor is sparse. The sparsity property is crucial for storage and computation speed.
	\item In the example studied in Section \ref{ssec:FungRoz} the grid spacing was larger than the noise. Since the grid must be able to sample the probability density, the effective noise vielbein was taken to be a constant ($1$ in our example) times the grid spacing, i.e., the signal model noise term is ``inflated''. Of course, this means that the result is not as accurate as the solution that uses the smaller values for the noise. However, it may still lead to acceptable results (as in the first example) at significantly lower computational effort.
	\item Observe that even with the coarse sampling the computed conditional probability density is ``smooth''. It seems apparent that a finer spatial grid spacing (with the same temporal grid spacing)will yield essentially the same result (using the DF approximation) at significantly higher computational cost. This was observed in the two examples studied in this paper. Of course, a multiple time step approximation would be more accurate. 
	\item Also note that the conditional mean estimation is quite good, i.e., of the order of the grid spacing, even for the coarser resolutions. This confirms the view that the conditional probability density calculated at grid points approximates very well the true value at those grid points (provided the computations are accurate). Alternatively, an interpolated version of the fundamental solution at coarser grid is close to the actual value. This suggests that a practical way of verifying the validity of the approximation is to note if the variation in the statistics with grid spacing, such as the conditional mean, is minimal.
	\item  It is also noted that the PDE-based methods are considerably more complicated for general two- or higher-dimensional problems. Specifically, the non-diagonal diffusion matrix case is no harder to tackle using path integral methods than the diagonal case. This is in sharp contrast to the PDE approach which for higher-dimensional problems is typically  based on operator splitting approaches. The operator splitting approaches cannot be reliable approximation for the general case.
	
	
	\item In this paper, the prediction and correction steps were carried out using a uniform grid. It is clear that a much faster approach for the correction part for higher dimensional problems would be to use Monte Carlo integration methods.
	\item Observe that the one-step approximation of the path integral can be stored more compactly. Compact representation of the transition probability density, especially in the It\^o case where it is of the Gaussian form. Even for the general case, the transition probability density from a certain initial point and given time step can be stored in terms of a few points with the rest obtainable via interpolation.
	\item Observe that the prediction step computation was sped up considerably by restricting calculation only in areas with significant conditional probability mass (or more accurately, in the union of the region of significant probability mass of $p(y(t_k)|x)$ and $p(x|Y (t_{k-1})))$. 
	\item It is noted that when the DF approximation is used with larger time steps, a coarser grid is more appropriate, which requires far fewer computations. Thus, a quasi-real-time implementation could use the coarse-grid approximation with larger time steps to identify local regions where the conditional pdf is significant so that a more accurate computation can then be carried out. 
	\item In this paper, it has been assumed that the diffusion matrix is invertible. Often this arises when it is a matter of defining a state variable as a time derivative of the other ($d\rv{x}_1(t) =\rv{x}_2(t)dt$). It is plausible that the addition of a small, positive perturbation to the diffusion matrix (so that the perturbed diffusion matrix is invertible) will not lead to large errors.
	\item When the step size is too large, the approximation will not be adequate. However, unlike some PDE discretization schemes, the degradation in performance is more graceful. For instance, positivity is always maintained since the transition probability density is manifestly positive. It is also significant to note that in physics path integral methods are used to compute quantities where $t'\ra-\infty$ and $t''\ra+\infty$ (see, for instance, \cite{IstvanMontvayandGernotMunster1997}). This is not possible by simple discretization of the corresponding PDE due to time step restrictions (note that implicit schemes are not as accurate).
	\item 	For the multiplicative noise case, the choice of $s\neq0$ leads to a more complicated form of the Lagrangian. The accuracy of the one-step approximation depends on $s$ in addition to $r$ and will be model-dependent.
	\item Note that, unlike the result of S-T. Yau and Stephen Yau in  \cite{S.-T.YauS.S.-TYau1996}, there is no rigorous bound on errors obtained for the Dirac-Feynman path integral formulae studied in the examples. It is known rigorously for a large class of problems that the continuum path integral formula converges to the correct solution\cite{R.AlickiD.Makoviec1985}.  

\end{enumerate}

\section{Conclusion}

In this paper, a new approach to solving the continuous-discrete filtering problem is presented. It is based on the Feynman path integral, which has been spectacularly successful in many areas of theoretical physics. The application of path integral methods to quantum field theory has also given striking insights to large areas of pure mathematics. The path integral methods has been shown offer deep insight into the solution of the continuous-discrete filtering problem that has potentially useful practical implications. In particular, it is demonstrated via non-trivial examples that the simplest approximations suggested by the path integral formulation can yield a very accurate solution of the filtering problem. The proposed Dirac-Feynman path integral filtering algorithm is very simple and easy to implement and practical for modest size problems, such as those arising in target tracking applications. Such formulae are also especially suitable from a real-time implementation point of view since it enables us to focus computation only on domains of significant probability mass. The application of path integral filtering for radar tracking problems, especially those with significant nonlinearity in the state model, will be investigated in subsequent papers. In a recent paper\cite{Balaji2008}, it has been shown that the Feynman path integral filtering techniques also leads to new insights into the general continuous-continuous nonlinear filtering problem. 

\section{Acknowledgements}
The author is grateful to Defence Research and Development Canada (DRDC) Ottawa for supporting this work under a Technology Investment Fund. 
\appendix

\section{Summary of Path Integral Formulas}\label{sec:AppendixI}
\subsection{Additive Noise}
The additive noise model
\begin{align}
	d\rv{x}(t)=f(\rv{x}(t),t)dt+e(t)d\rv{v}(t),
\end{align}
is interpreted as the continuum limit of
\begin{align}
	\Delta \rv{x}(t)=f(\rv{x}^{(r)}(t),t)\Delta t+e(t)\Delta\rv{v}(t), 
\end{align}
where
\begin{align}
	\rv{x}^{(r)}(t)=\rv{x}(t-\Delta t)+r(\rv{x}(t)-\rv{x}(t-\Delta t)).
\end{align}
Observe that any $r\in[0,1]$ leads to the same continuum expression.

The transition probability density for the additive noise case is given by (see \cite{PAPER1})
\begin{align}
	P(t'',x''|t',x')=\int_{x(t')=x'}^{x(t'')=x''}[\mc{D}x(t)]\exp\left(-\int_{t'}^{t''}dtL^{(r)}(t,x,\dot{x}) \right),	
\end{align}
where the Lagrangian $L^{(r)}(t,x,\dot{x})$ is
\begin{align}
	L^{(r)}(t,x,\dot{x})=\frac{1}{2}\sum_{i=1}^n\left[ \dot{x}_i-f_i(x^{(r)}(t),t) \right]g^{-1}_{ij}(t)\left[ \dot{x}_j-f_j(x^{(r)}(t),t) \right]+r\sum_{i=1}^n\frac{\partial f_i}{\partial x_i}(x^{(r)}(t),t),
\end{align}
and $g_{ij}(t)=\sum_{a,b=1}^{p_e}e_{ia}(t)Q_{ab}(t)e_{jb}(t)$, and
\begin{align}
	\left[ \mc{D}x(t) \right]=\frac{1}{\sqrt{(2\pi\epsilon)^n\det g(t')}}\lim_{N\ra\infty}\prod_{k=1}^N\frac{d^nx(t'+k\epsilon)}{\sqrt{(2\pi\epsilon)^n\det g(t'+k\epsilon)}}.
\end{align}
This formal path integral expression is defined as the continuum limit of\
\begin{align}
	\frac{1}{\sqrt{(2\pi\epsilon)^n\det g(t'')}}\int\prod_{k=1}^N\left[ d^nx(t'+k\epsilon) \frac{1}{\sqrt{(2\pi\epsilon)^n\det g(t'+k\epsilon)}}\right]\exp\left( -S_{\epsilon}^{(r)}(t'',t') \right),
\end{align}
where the discretized action $S_{\epsilon}^{(r)}(t'',t')$ is defined as
\begin{align}
	&\frac{1}{2\epsilon}\sum_{k=1}^{N+1}\left[ \sum_{i,j=1}^n(x_i(t_k)-x_i(t_{k-1})-\epsilon f_i(x^{(r)}(t_k),t_k))g_{ij}^{-1}(x_j(t_k)-x_j(t_{k-1}+\epsilon f_j(x^{(r)}(t_k),t_k))) \right]\\ \nonumber
	&\qquad+\sum_{k=1}^{N+1}\left[ r\sum_{i=1}^n\frac{\partial f_i}{\partial x_i}(x^{(r)}(t_k),t_k) \right],
\end{align}
and where
\begin{align}
	x^{(r)}(t_k)=x(t_{k-1})+r(x(t_k)-x(t_{k-1})).
\end{align}

\subsection{Multiplicative Noise}
Consider the evolution of the stochastic process in the time interval $[t', t'']$. Divide the time interval into $N + 1$ equi-spaced points and define $\epsilon$ by $t' + (N + 1)\epsilon = t''$, or $\epsilon=\frac{t''-t'}{N+1}$. Then, in discrete-time, the most general discretization of the Langevin equation is
\begin{align}\label{eq:DiscLangGen00}	
\rv{x}_i(t_p)-\rv{x}_i(t_{p-1})=\epsilon f_i(\rv{x}^{(r)}(t_p),t_p)+e_{ia}(\rv{x}^{(s)}(t_p),t_p)(\rv{v}_a(t_p)-\rv{v}_a(t_{p-1})),
\end{align}
where $p=1,2,\ldots,N+1, 0\le r,s\le1$, and
\begin{align}
        \rv{x}_i^{(r)}(t_{p})&=\rv{x}_i(t_{p-1})+r\Delta\rv{x}_i(t_p),&\rv{x}_i^{(s)}(t_{p})&=\rv{x}_i(t
_{p-1})+s\Delta\rv{x}_i(t_p),\\ \nonumber
        &=\rv{x}_i(t_{p-1})+r(\rv{x}_i(t_p)-\rv{x}_i(t_{p-1})),&&=\rv{x}_i(t_{p-1})+s(\rv{x}_i(t_p)-\rv{
x}_i(t_{p-1})).
\end{align}
In this section, the Einstein summation convention is adopted, i.e., all repeated indices are assumed to be summed over, so that $e_{ia}d\rv{v}_a=\sum_{a=1}^pe_{ia}d\rv{v}_a$. Also, $\frac{\partial}{\partial x_i^{(r)}}$ is written as $\partial_i^{(r)}$. 

Note that the change in Equation \ref{eq:DiscLangGen00} when $f_i(\rv{x}^{(r)}(t_{p}),t_{p})$ and $e_{ia}(\rv{x}^{(s)}(t_{p}),t_{p})$ are replaced with $f_i(\rv{x}^{(r)}(t_{p}),t_{p-1})$ and $e_{ia}(\rv{x}^{(s)}(t_{p}),t_{p-1})$ is of $O(\epsilon^2)$ and $O(\epsilon^{3/2})$ respectively. Hence, it may be ignored in the continuum limit as it is of order higher than $O(\epsilon)$. 

In summary, there are infinitely many possible discretizations parametrized by two reals $r, s\in[0, 1]$. In the continuum limit, i.e., $\epsilon\ra0$, observe that the stochastic process depends on $s$, but not on $r$. When $s = 0$, the limiting equation is said to be interpreted as an It¿o SDE, while when $s = \frac{1}{2}$ , the equation is said to be interpreted in the Stratanovich sense.

In \cite{PAPER3} it is shown that for the general multiplicative noise case
\begin{align}
	P(t'',x''|t',x')=\int_{x(t')=x'}^{x(t'')=x''}\left[ \mc{D}x(t) \right]\exp\left( -S^{9r,s)} \right),
\end{align}
where the action $S^{(r,s)}$ is given by
\begin{align}
	S^{(r,s)}=\int_{t'}^{t''}dt\left[ \frac{1}{2}J_i^{(r,s)}\left( g^{-1} \right)_{ij}J_j^{(r,s)}+r\partial_i^{(r)}f_i+\frac{s^2}{2}\left[ (\partial_i^{(s)}e_{ja})Q_{ab}(t_p)(\partial_j^{(s)})e_{ia} -(\partial_i^{(s)}e_{ia})Q_{ab(t)}(\partial_j^{(s)}e_{ja})\right] \right],
\end{align}
where
\begin{align}
	g_{ij}=\sum_{a,b=1}^{p_e}e_{ia}(x^{(s)}(t),t)Q_{ab}(t)e_{jb}(x^{(s)}(t),t),
\end{align}
and
\begin{align}
	J_i^{(r,s)}=\left( \frac{dx_i}{dt}(t)-f_i(x^{(r)}(t),t)-s\sum_{a,b=1}^{p_e}\sum_{i'=1}^{n}e_{ia}(x^{(s)}(t),t)Q_{ab}(t)\frac{\partial e_{i'b}}{\partial x_{i'}^{(s)}}(x^{(s)}(t),t) \right),
\end{align}
and the probability measure $[\mc{D}x(t)]$ is given by
\begin{align}
	\frac{1}{\sqrt{(2\pi\epsilon)^n\det g(x^{(s)}(t''),t'')}}\left[ \prod_{p=1}^{N}\left\{ \frac{d^nx(t_p)}{\sqrt{(2\pi\epsilon)^n\det g(x^{(s)}(t_p),t_p)}} \right\} \right].
\end{align}
The discretized expression for the general case is complicated but can be written down from these results in a straightforward manner.

\bibliographystyle{IEEEtran.bst}
\bibliography{onfbib}

\end{document}